\documentclass[usenatbib]{mnras}

\usepackage{newtxtext,newtxmath}
\usepackage{xspace,amsmath}
\usepackage{color,epsfig,amssymb,lscape}
\usepackage{array,colortbl}
\usepackage{url}

\usepackage[T1]{fontenc}
\usepackage{ae,aecompl}

\newcommand{\hii}{\relax \ifmmode {\mbox H\,{\scshape ii}}\else H\,{\scshape ii}\fi}
\newcommand{\mi}{\relax \ifmmode {\mu{\mbox m}}\else $\mu$m\fi}
\newcommand{\ha}{\relax \ifmmode {\mbox H}\alpha\else H$\alpha$\fi}
\newcommand{\hb}{\relax \ifmmode {\mbox H}\beta\else H$\beta$\fi}

\newcommand{\sii}{\relax \ifmmode {\mbox S\,{\scshape ii}}\else S\,{\scshape ii}\fi}
\newcommand{\siii}{\relax \ifmmode {\mbox S\,{\scshape iii}}\else S\,{\scshape iii}\fi}
\newcommand{\nii}{\relax \ifmmode {\mbox N\,{\scshape ii}}\else N\,{\scshape ii}\fi}
\newcommand{\oi}{\relax \ifmmode {\mbox O\,{\scshape i}}\else O\,{\scshape i}\fi}
\newcommand{\oii}{\relax \ifmmode {\mbox O\,{\scshape ii}}\else O\,{\scshape ii}\fi}
\newcommand{\oiii}{\relax \ifmmode {\mbox O\,{\scshape iii}}\else O\,{\scshape iii}\fi}
\newcommand{\neiii}{\relax \ifmmode {\mbox Ne\,{\scshape iii}}\else Ne\,{\scshape iii}\fi}

\newcommand{\rdostres}{\relax \ifmmode {\,\mbox{R}}_{\rm 23}\else \,\mbox{R}$_{\rm 23}$\fi} 

\newcommand{\ciii}{\relax \ifmmode {\mbox O\,{\scshape iii}}\else C\,{\scshape iii}\fi}
\newcommand{\civ}{\relax \ifmmode {\mbox O\,{\scshape iii}}\else C\,{\scshape iv}\fi}
\newcommand{\heii}{\relax \ifmmode {\mbox He\,{\scshape ii}}\else He\,{\scshape ii}\fi}

\newcommand{\refpmv}{P\'erez-Montero \& V\'\i lchez (2009)}

\newcommand{\gsim}{\hbox{\rlap{\lower.55ex\hbox{$\sim$}} \kern-.3em
\raise.4ex \hbox{$>$}}}
\newcommand{\lsim}{\hbox{\rlap{\lower.55ex\hbox{$\sim$}} \kern-.3em
\raise.4ex \hbox{$<$}}}

\usepackage{graphicx}	
\usepackage{amsmath}	
\usepackage{amssymb}	

\title[SED hardening in galaxy discs]{Revisiting the hardening of the
stellar ionizing radiation in galaxy discs}

\author[Enrique P{\'e}rez-Montero et al.]{
Enrique P{\'e}rez-Montero,$^{1}$\thanks{E-mail: epm@iaa.es (EPM)},
Rub\'en Garc\'\i a-Benito$^1$ \& Jos\'e M. V\'\i lchez$^1$ 
\\
$^{1}$Instituto de Astrof\'\i sica de Andaluc\'\i a. CSIC. Apartado de correos 3004. 18080, Granada, Spain.\\
}

\date{Accepted XXX. Received YYY; in original form ZZZ}

\pubyear{2017}

\begin{document}
\label{firstpage}
\pagerange{\pageref{firstpage}--\pageref{lastpage}}
\maketitle

\begin{abstract}
In this work we explore accurate new ways to derive the ionization parameter ($U$) 
and the equivalent effective temperature ($T_*$) in \hii\ regions using emission-line intensities
from the ionized gas. The so-called softness parameter ($\eta$), based on [\oii], [\oiii], [\sii], and [\siii] has
been proposed to estimate the hardening of the ionizing incident field of radiation, but the simplest 
relation of this parameter with $T_*$ also depends on $U$ and metallicity ($Z$). Here we provide a Bayesian-like code
({\sc HCm-Teff}) that compares the observed emission lines of $\eta$ with the
predictions of a large grid of photoionization models giving precise estimations of both $U$ and $T_*$
when $Z$ is known.
We also study the radial variation of these parameters in 
well-studied disc galaxies observed by the {\sc chaos} collaboration. 
Our results indicate that the observed radial decreasing of $\eta$ can be attributed
to a radial hardening of $T_*$, across galactic discs
as in NGC~628 and NGC~5457. On the other hand NGC~5194, which presents a
positive slope of the fitting of the softness parameter, has a flat slope in $T_*$.
On the contrary the three galaxies do not seem to
present large radial variations of the ionization parameter.
When we inspect a larger sample of galaxies 
we observe steeper radial variations of
$T_*$ in less bright and later-type  galaxies, mimicking a similar trend observed for $Z$
but the studied sample should be enlarged to obtain more statistically significant 
conclusions.

\end{abstract}

\begin{keywords}
methods: data analysis -- ISM: abundances -- galaxies: abundances
\end{keywords}



\section{Introduction}

Objects whose luminosity is dominated by conspicuous emission lines from ionized gas
are preferential targets in galaxies for the determination of many different physical
properties of the spatial regions where they are located.
Though the weighted average of the ionic species heavier than helium is usually
lower than 2\%, the relative intensity of the optical lines emitted by these 
elements is easily detectable in bright star-forming complexes. In the case of those
elements that present more than one collisionally excited emission line corresponding to
different ionization stages in the same spectral observed range,
their study allows us to extract important conclusions about the photoionization 
equilibrium between the hardening of the ionizing radiation from massive stars and
the density of elements in the surrounding gas. 

According to \cite{mrs85} most of the
properties of the integrated emitting spectrum depend on three properties or so-called
functional parameters: metallicity ($Z$), ionization parameter ($U$), and 
equivalent effective temperature ($T_*$). 
$Z$ is usually derived by means of measurement of the
electron temperature using collisionally excited emission lines from one ion 
whose energies are similar (e.g. [\oiii] 5007/4363). Alternatively, when no auroral emission
line can be measured with enough confidence in the emission-line spectrum, we can resort to 
direct calibrations of the strong emission lines. $U$ can be estimated using
ratios of emission lines corresponding to consecutive ionization stages, such as [\oii]/[\oiii] and [\sii]/[\siii],
although these ratios usually also present a strong additional dependence on the other functional parameters (e.g. \citealt{diaz98,dors03}).

To calculate $T_*$ several authors have proposed different
sensitive emission-line ratios, such as [Ne{\sc iii}] 3869 \AA/\hb\ \citep{oey2000} 
and  He{\sc i}/\hb\ \citep{asari16}.
\cite{vp88} proposed the use
of two different ionic abundance ratios of consecutive ionization stages corresponding to
species with very different excitation energies to trace the shape of the spectral ionizing energy distribution (SED).
In the case of the optical range this can be obtained using oxygen and sulphur abundances that define the so-called softness parameter:

\begin{equation}
\eta = \frac{O^+/O^{2+}}{S^+/S^{2+}}
\end{equation}

The softness parameter decreases at high $T_*$ of the
field of radiation while it increases for low values of $T_*$.

These authors also defined the corresponding electron-temperature-independent line ratio quotient as:

\begin{equation}
\eta\prime = \frac{[OII] 3727/[OIII] 4959,5007 }{[SII] 6717,6731 / [SIII] 9059,9532}
\end{equation}

Other authors have also proposed other sets of lines both in the optical 
(e.g. [Ar{\sc iii}]/[Ar{\sc iv}]\ vs [\oii]/[\oiii]; \citealt{stasinska15}) 
and in the mid-infrared (e.g. [Ne{\sc ii}]/[Ne{\sc iii}]\ vs [S{\sc iii}]/[S{\sc iv}]; \citealt{morisset04,pmv09}).

The use of this variant of the softness parameter based only on emission lines
 presents additional dependence on both $Z$ and $U$
(e.g. \citealt{morisset04}). In addition a direct comparison between
log $\eta\prime$ and $T_*$ of the field
of radiation can result in uncertainty  because the curves of equal $T_*$
predicted by single-star photoionization models
have slopes lower than 1 in the plane
log([\sii]/[\siii]) vs log([\oii]/[\oiii]), so a spatial variation of $\eta\prime$
can also involve a variation of $U$ (e.g. \citealt{sca11,fm17}). 
In this  way, \cite{pm14} show in a 3D modelling of the giant \hii\ region NGC~595 that it can be 
predicted by a radial variation of log $\eta\prime$ only owing to the decrease of $U$
even if the ionizing central source is the same for all layers of the ionized gas.
Therefore, it is necessary to provide new and more accurate 
recipes to give a trustable estimation of $T_*$ using the available optical
emission lines.

One of the open issues on which a thorough study of the spatial variation of $T_*$  can
shed some light is the radial variation of the functional parameters across spiral discs, 
which would help to constrain models of formation and build-up
of these structures along cosmic time.
From \cite{searle71}, it is known that there is a general trend to find higher
excitations of the gas at larger galactocentric distances. This was soon linked with the
presence of a decreasing radial variation of $Z$ (\citealt{smith75}).
The slope of these gradients correlates with the mass and luminosity of galaxies (e.g. \citealt{garnett97,pvc04}).
The resulting slopes of the obtained fittings can even be normalized by the effective radius \citep{diaz89}, 
which can ease the study of the global radial variation of properties such as the total oxygen abundance (e.g. \citealt{sanchez14})
and the nitrogen-to-oxygen ratio \citep{pm16}.

From the point of view of radial variation across galactic discs of $T_*$,
\cite{ss78} were the first to propose a radial enhancement in NGC~5457 based on the higher equivalent widths of \hb\
observed in the outermost \hii\ regions. Several authors have confirmed this trend in other spiral galaxies by means of a direct
comparison between emission-line fluxes and photoionization models (e.g. \citealt{fierro86, dors05, dors17}. On the other hand this radial variation
of $T_*$ with galactocentric distance is not apparently appreciated in the Milky Way \citep{morisset04}.

\cite{pmv09} also measured a radial variation of
the softness parameter in different disc galaxies. Most of the studied objects
present lower values of $\eta\prime$ at larger galactocentric distances, which could
be interpreted as a hardening of the ionizing stellar radiation across galactic discs.
As in the case of $Z$ the scale of this variation seems to be related to
the size and mass of the studied galaxies, though the sample of studied objects is still
quite limited to establish solid conclusions in this respect.
Therefore applying a new thorough analysis to a more complete sample of \hii\ regions in
well-characterized galaxies would supply some clarification to the existence of
gradients of one or various of the functional parameters in discs
along with a possible connection between them and with other integrated properties of galaxies,
as in the case of metallicity.

In this paper we present a new routine based
on photoionization models to break the degeneracy of the softness parameter
when it is calculated from emission lines
and hence to provide precise estimates of both $U$ and $T_*$ in regions whose metallicity has been previously very well defined
and with accurate measurements of the four optical emission lines involved in the
calculation of $\eta\prime$.
Our objective is to apply it to a sample of galaxies to study the radial variation of the derived properties
and to verify whether the observed radial variation of the {\em softness} parameter can be attributed to a real 
variation of the hardening of the incident field of radiation across 
galactic discs.

The paper is organized as follows. In Section 2 we describe the new routine to obtain $U$ and $T_*$ using
the emission lines involved in the definition of the softness parameter
with the help of a large grid of photoionization models.
Section 3 is devoted to comparing the results from this code with other models 
for both single stars and star clusters and with observational data.
In Section 4 we study the radial variation of the softness parameter and the derived
$U$ and $T_*$ using the code described in the previous sections in 
a sample of three galaxies of the {\sc chaos} (CHemical Abundances Of
Spiral galaxies) project \cite{chaos1}) with good determinations of the chemical abundances
following the direct method in a large number of \hii\ regions across the galactic discs.
In Section 5 we compare the results obtained for the three {\sc chaos} galaxies with the objects analysed in \cite{pmv09}
and we discuss the relation between the obtained slopes with other integrated properties
in the same galaxies.
Finally,  in Section 6 we summarize our results and we present
our conclusions.

\begin{figure}
\centering

\includegraphics[width=8cm,clip=]{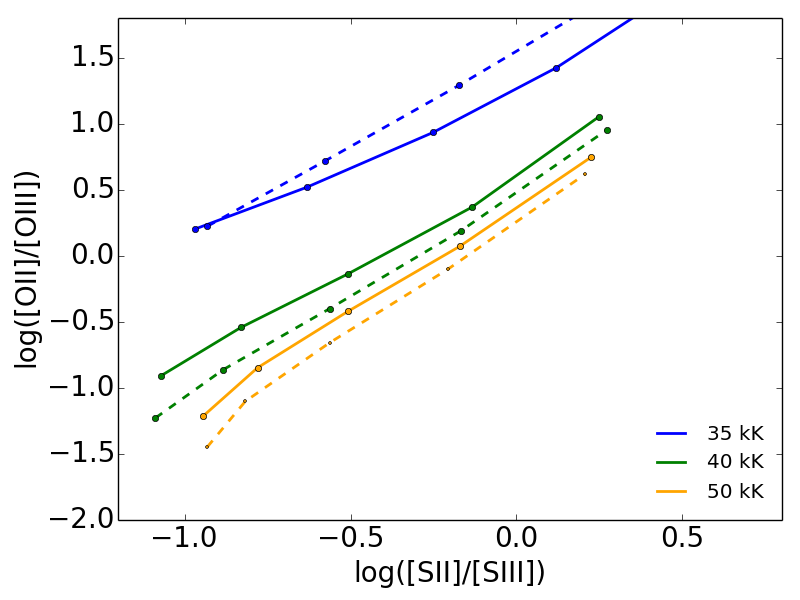}

\caption{Relations between the emission-line ratios [\sii]/[\siii] and
[\oii]/[\oiii] predicted by some of the models described in the text. Each color 
represents different values of $T_*$.
The values of log $U$ go from -3.5 to -1.5 in intervals of 0.5 dex from the lower left to
the upper right part of the diagram.
 The solid lines join models with $Z$ = $Z_{\odot}$ while
the dashed lines join models with $Z$ = $Z_{\odot}$/5.}

\label{eta_models}
\end{figure}

\section{Model and method description}

The routine designed to calculate both $U$ and 
$T_*$ from the emission lines used in the definition of the $\eta\prime$
parameter is very similar to the {\sc Hii-Chi-mistry} \citep{hcm} code used to
derive chemical abundances in nebulae ionized by massive stars.
The method consists of comparison between certain observed emission-line fluxes and the corresponding
predictions made by a large grid of photoionization models covering the expected 
conditions in the studied object or position in an object.

\begin{figure*}
\centering

\includegraphics[width=8cm,clip=]{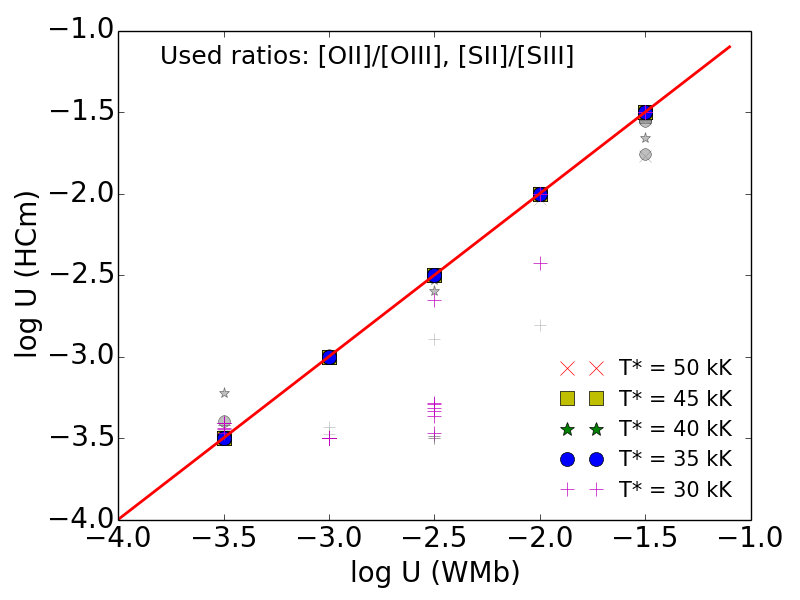}
\includegraphics[width=8cm,clip=]{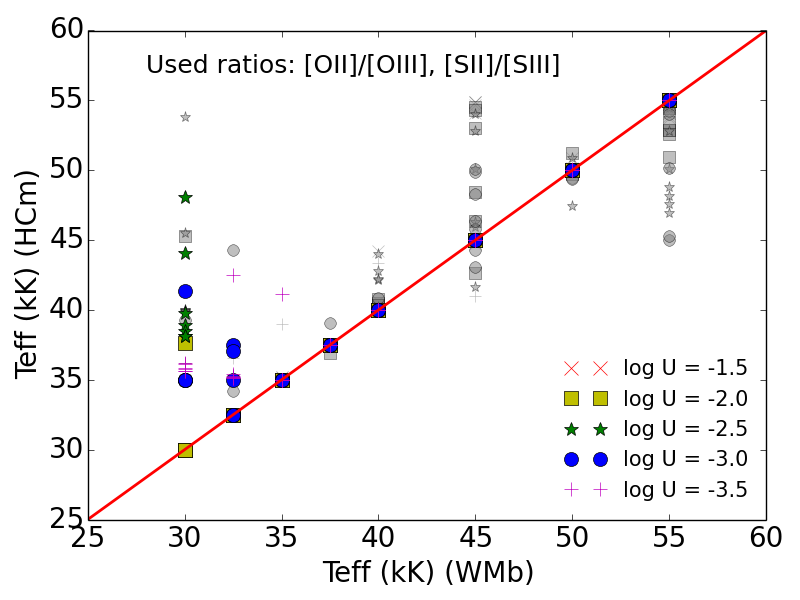}
\includegraphics[width=8cm,clip=]{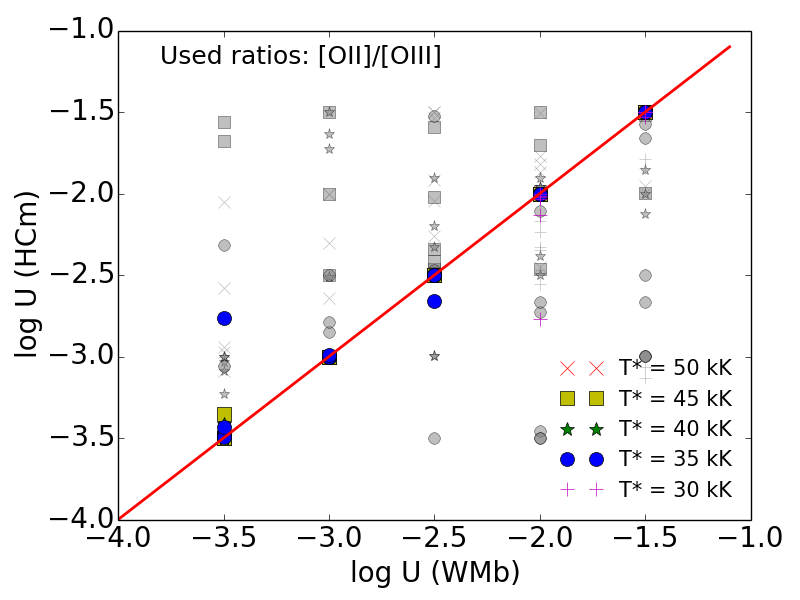}
\includegraphics[width=8cm,clip=]{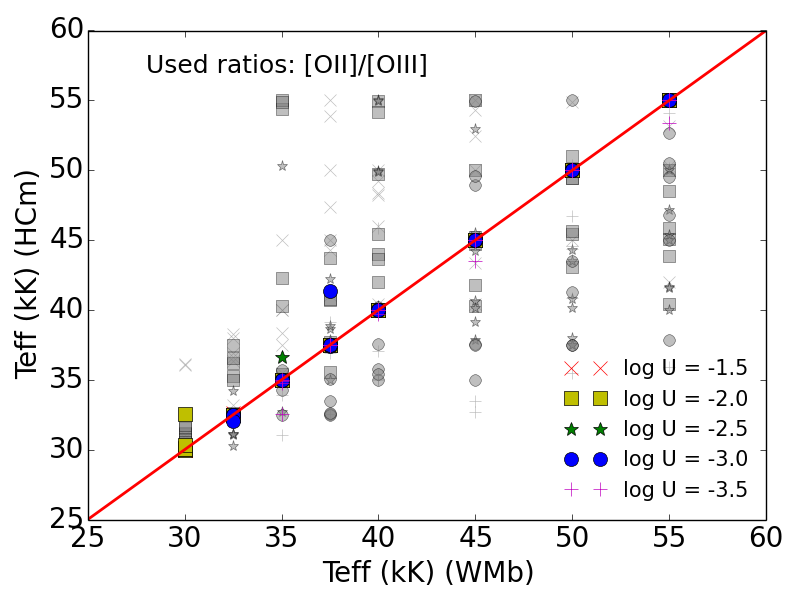}
\includegraphics[width=8cm,clip=]{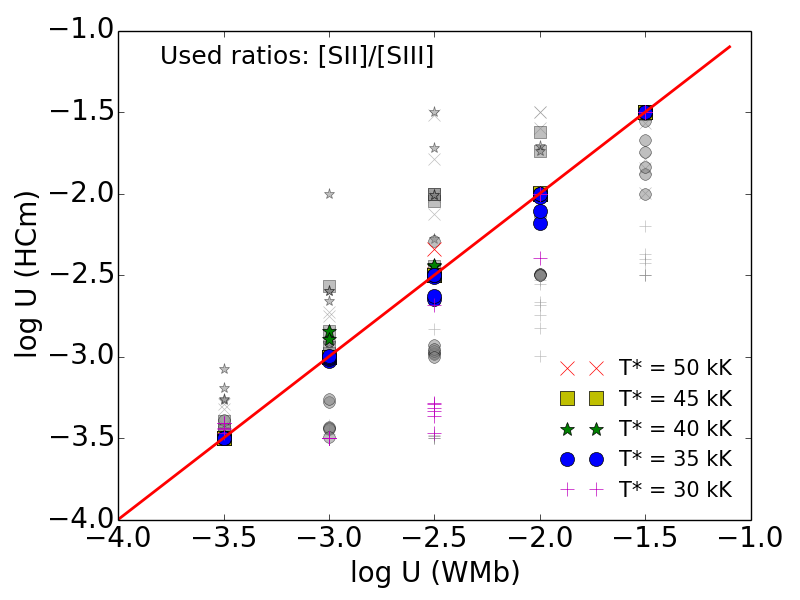}
\includegraphics[width=8cm,clip=]{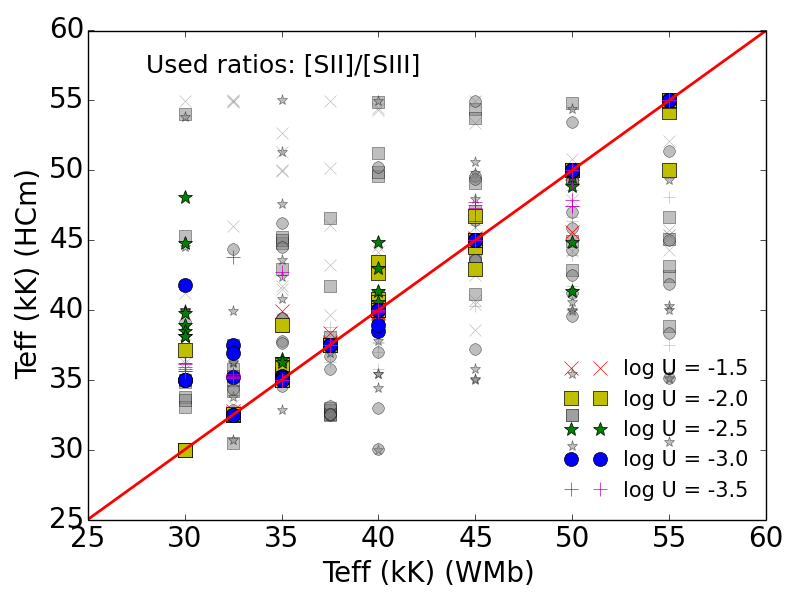}

\caption{Comparison between the input $U$ (left-hand column) and $T_*$ (right-hand column) of the grid of models
and the same properties predicted by our code using the corresponding
different sets of  emission-line ratios. 
In the top row the code uses  as input both [\oii]/[\oiii] and [\sii]/[\siii]
line ratios. In the middle row we show the comparison when only the [\oii]/[\oiii] emission-line ratio is used. Finally, in the bottom row we show the results when the code only makes use of the [\sii]/[\siii] emission-line ratio.
In all panels thick coloured symbols correspond to calculations performed using metallicity as an input of the code,
while thin-grey symbols leave metallicity as a free parameter.}

\label{comp}

\end{figure*}

The grid of models used in the routine described here was calculated using the code {\sc Cloudy} v.17.00
\citep{cloudy}, which gives the relative intensities of the lines emitted by a 
one-dimensional distribution of gas ionized by a central source. For each model single ionizing source 
we used the SEDs of WM-Basic \citep{wmbasic} stellar atmospheres 
with $T_*$ values at 30, 32.5, 35, 37.5, 40, 45, 50, and 55 kK at the same $Z$
as the surrounding gas distribution. 
According to \cite{zastrow13} the synthetic WM-Basic SEDs in photoionization
models lead to a better agreement with observed emission-line ratios in 
single-star \hii\ regions where they can
be directly compared with the SED of the ionizing source.

For the gas we considered 
different $Z$ values scaled to the one of the oxygen abundance using the solar proportions 
given by \cite{asplund09} at the values 12+log(O/H) = 7.1, 7.4, 7.7, 8.0,
8.3, 8.6, and 8.9. We assumed in all models a constant electron density of
100 particles cm$^{-3}$ and a standard gas-to-dust mass ratio
and depletion of the corresponding refractory elements.
For $U$ we covered the values
log $U$ = -3.5, -3.0, -2.5, -2.0, and -1.5. The stopping criterion to recover the calculated set
of emission lines was reached when the electron temperature of the gas was lower than 4\,000 K,
which results in all models in a spherical geometry.
The total number of models of the grid is then 280.

The python code {\sc Hii-Chi-mistry-Teff} \footnote{Publicly available in the
webpage \url{http://www.iaa.csic.es/~epm/HII-CHI-mistry-Teff.html}} (hereafter {\sc HCm-Teff}) makes a direct comparison between the
model-predicted [\oii] 3727 \AA\ relative to [\oiii] 5007 \AA\ and [\sii] 6717,6731 \AA\ 
relative to [\siii] 9069 \AA\ emission-line ratios with the corresponding reddening-corrected observed values in arbitrary units
to give an estimation of both $U$ and T$_*$. 
Although in the definition of the $\eta\prime$ parameter both lines of the  
[\oiii] and [\siii] doublets are included, as there is a theoretical
relation between the intensities of these lines [i.e. I(5007 \AA) = 3$\times$ I(4959) 
and I(9069) = 2.44 $\times$I(9532 \AA]
that does not make it necessary 
to use of all them. In the case of [\siii] this is convenient as many times the emission line at 9532 \AA\ is affected by
atmospheric absorption, although this also depends on the redshift of the studied object. 
In addition it is necessary to be cautious with the extinction correction of these
lines in the near-IR when it is performed in relation to \hb\ and it is convenient to use the close hydrogen recombination lines of the Paschen series
to provide relative intensities that can be later expressed in terms
of other lines in the optical range.

The code requires the oxygen abundance in
each input observed object or position as the models predict a large dependence
of the emission lines on $Z$.
In Figure \ref{eta_models} we see the relation between both oxygen and sulphur
optical emission-line ratios and the prediction from our grid of models for different $T_*$
at two $Z$ values ($Z_{\odot}$ and $Z_{\odot}/5$).  As can be seen, 
the lines of equal $T_*$ tend to lie in a lower position for lower $Z$ and this effect
is more noticeable for higher $T_*$. For this reason the code first interpolates a subset of
emission-line predictions at the chosen oxygen abundance. In case this
parameter is not provided the code assumes the complete grid of models in the
calculation, which translates to a larger uncertainty of the final derived products.

Once the grid of models is chosen according to the provided $Z$, 
the code calculates both $U$ and $T_*$  as the weighted mean in the distribution of all
the values in each model $i$ in the grid as

\begin{equation}
\log(U)_f = \frac{\sum_i \log(U)_i/\chi^2_i}{\sum_i 1/ \chi^2_i}
\end{equation}

\noindent and

\begin{equation}
T_{*f} = \frac{\sum_i T_{*i}/\chi^2_i}{\sum_i 1/ \chi^2_i}
\end{equation}

\noindent where the $\chi_i$ values in each model are calculated as
the sum over the set of the normalized difference between certain
observables, $O_j$, and the corresponding values predicted by each model, $T_j$, as

\begin{equation}
\chi_i^2 =\sum_j \frac{(O_j - T_{ji})^2}{O_j}
\end{equation}

The emission-line ratios used by the code as observed quantities  are log ([\oii]/[\oiii]) and log([\sii]/[\siii])
In case the four required lines are not available the code also allows us to perform the calculations using only one of
the above emission-line ratios.

The errors are calculated as the standard deviations of these 1/$\chi^2$-weighted distributions.
The code also requires the observed errors in each line,
if available,  and calculates their contribution to the  
corresponding $U$ and $T_*$  errors using a Monte Carlo
iteration randomly varying the input intensities  around the provided input errors.
Then these two error sources are quadratically added by the code.

\section{Comparison with models and observations}

In order to test the validity of our method to estimate both $U$ and $T_*$
using the  emission lines in the softness parameter we took the predicted emission lines
in the same WM-Basic single-star models as input for our code and we checked whether we recovered the same input $U$ and $T_*$.

In Figure \ref{comp} we show a comparison between the conditions imposed in each model
and the log $U$ and $T_*$ values derived from the {\sc HCm-Teff} code using only the lines.
We also checked how using metallicity as an input in the code impacts the accuracy of the results.
We tested three different cases attending to the use of both oxygen and sulphur emission-line ratios or using only one of them.

As can be seen we
recover values close to the conditions of the models
for both $U$ and $T_*$  when we use the two emission-line ratios
using the metallicity as an input.
For log $U$ the mean offset is lower than 0.05 dex at all values 
with a mean standard deviation of the residuals of 0.2 dex
while for $T_*$ for values larger than 35kK the mean
offset and the standard deviation of the residuals are both lower than 500 K.
When metallicity is not added as an input to the code, but it is left as a free parameter and the code
has to explore the entire grid, both $T_*$ and $U$
become more uncertain.
Anyway, for $T_*$ the standard deviation of the residuals is on average 2500 K and the mean offset is always lower than 5000 K.
On the contrary, for $U$ the dependence on metallicity looks to be lower, as the mean offset is
lower in all cases than 0.1 dex, although the uncertainty is higher for lower excitations.

When we use only one of the two involved emission-line ratios we clearly see that
[\oii]/[\oiii] is more appropriate for the calculation of $T_*$ while [\sii]/[\siii] gives reliable estimates of
log $U$. The offsets and standard deviations between
the model conditions and those derived by our code are very similar to those obtained using both
emission-line ratios in these quantities. On the other hand
[\oii]/[\oiii]  can lead to deviations larger than 1 dex for low values of log $U$
and [\sii]/[\siii] 
can overestimate $T_*$ on average more than 5 kK at values lower than 40Kk.
This trend is even more pronounced when metallicity is not used as an input
in the code and the derivation of T$_*$ using only [\oii]/[\oiii] and log $U$ from [\sii]/[\siii] is much
worse than when we have the two emission-line ratios.
Besides, the determination of $U$ using [\oii]/[\oiii] 
has an uncertainty higher than 0.5 dex in all cases, and even larger than 1 dex for low excitation.
In the case of [\sii]/[\siii], the uncertainty to derive $T_*$ is never lower than 5000 K.

\begin{figure}
\centering

\includegraphics[width=8cm,clip=]{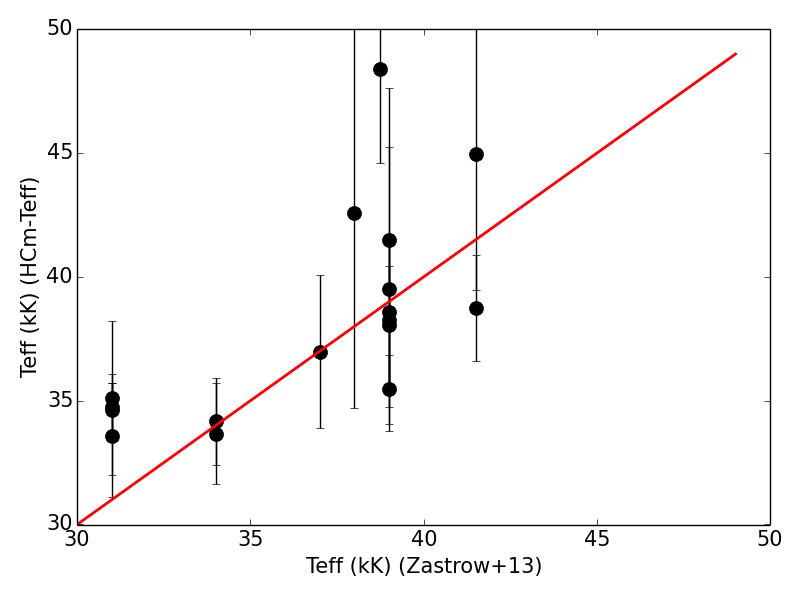}

\caption{Comparison between the derived $T_*$ 
in single-star \hii\ regions given by Zastrow et al. (2013) and the results obtained from our code
using only [\oii] and [\oiii] emission lines, as provided by the authors.}

\label{comp_Zastrow}
\end{figure}

\begin{figure*}
\centering

\includegraphics[width=8cm,clip=]{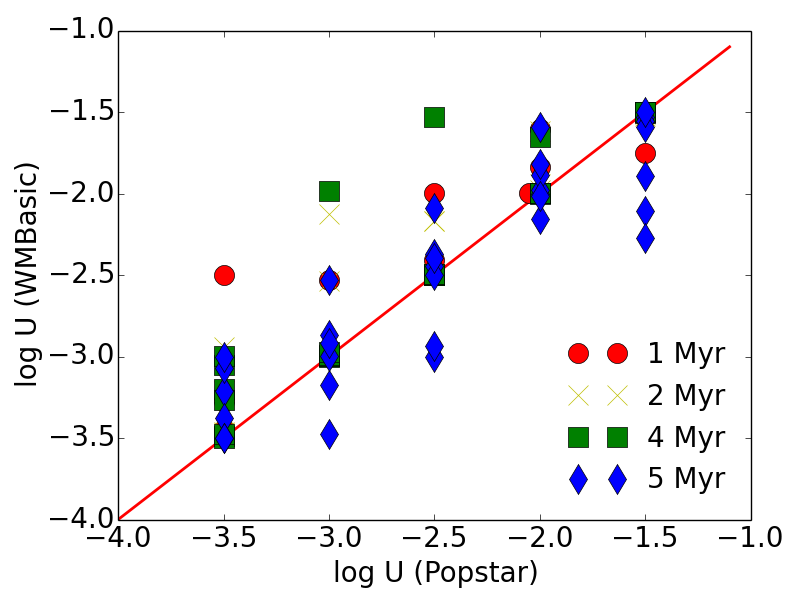}
\includegraphics[width=8cm,clip=]{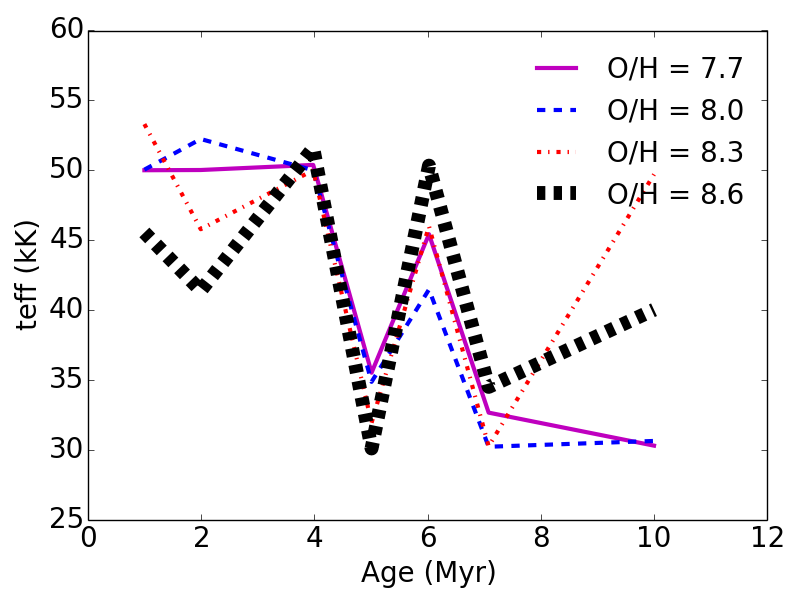}

\caption{Results from the analysis of emission lines from
models with {\sc popstar} synthetic cluster SEDs using {\sc HCm-Teff}.
Left: comparison between the $U$ used as input in the models
and the values predicted by the code for different ages.
Right: relation between the age of the cluster
and $T_*$ predicted by our code using the emission lines
for different $Z$ values.
} 

\label{popstar_Teff}
\end{figure*}

In addition it is possible to establish a direct comparison with data taking the spectroscopic optical observations
of single-star \hii\ regions of the Galaxy given by \cite{zastrow13}. These authors provide fittings
from WM-Basic SED synthetic atmospheres and flux measurements of the most prominent emission lines in different long-sit positions in the surrounding gas,
although the [\sii]/[\siii] emission-line ratio is not provided in this work.
In Figure \ref{comp_Zastrow} we show the comparison between the $T_*$ values 
given by these authors and the values derived by our code using only 
[\oii] and [\oiii], with the addition of $Z$ in those pointings with
a measurement of the [\oiii] auroral emission line at 4363 \AA. 
As can be seen the scale of temperatures of the ionizing star is relatively well reproduced 
and the obtained values agree within the errors with the estimations given by the authors.

As a final test we checked the effect of applying our code using results from photoionization models based on single-star SEDS to
the results from models using SEDs from massive clusters. To this aim we also
produced a grid of photoionization models with the same gas conditions as described above but using a grid of {\sc POPSTAR} \citep{popstar} with the same $Z$ as those
defined in the grid of the code but with ages of 1,2,4,5, 7 and 10 Myr. 
We then used the resulting [\oii], [\oiii], [\sii], and [\siii] 
predicted emission lines with their corresponding input $Z$ as input in our code.

In the left-hand panel of Figure \ref{popstar_Teff}  we show a comparison between the values of log $U$s chosen as input
in the photoionization models calculated in this work using the {\sc popstar}
cluster synthetic SEDs and the results provided by {\sc HCm-Teff} using the emission lines
as input for the code. As can be seen the
agreement is better for younger ages of the cluster. For an age of 1 Myr the mean offset of the log $U$ values is lower than 0.01 dex with
a standard deviation of the residuals of only 0.1 dex. This dispersion grows with age and it is around 0.5 dex for
ages larger than 5 Myr.

In the right-hand panel of the same figure we show a comparison between the resulting $T_*$ predicted by our 
code using only the emission lines and the corresponding cluster ages of the input models. As expected $T_*$ is higher for
younger ages, with a decrease with age, a stage of hardening of the radiation corresponding to the Wolf-Rayet phase 
and a final decrease for old ages. 
The derived $T_*$ is lower at a specific age when $Z$ enhances as is expected from the effect of a larger blanketing 
from metals in the star atmospheres. Even for 12+log(O/H) = 8.6, the photoionization models do not
predict the emission of high-excitation lines like [\oiii] and [\siii] for ages larger than 6 Myr.

\begin{figure}
\centering

\includegraphics[width=8cm,clip=]{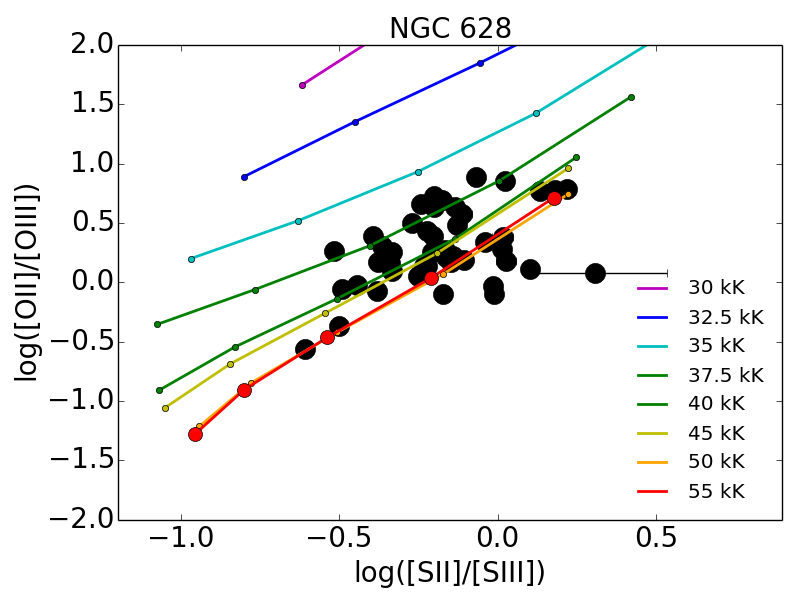}
\includegraphics[width=8cm,clip=]{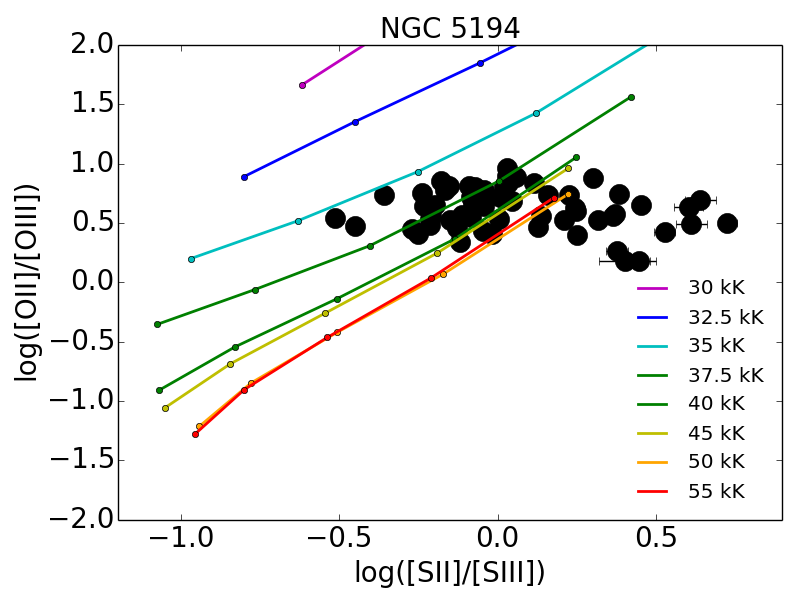}
\includegraphics[width=8cm,clip=]{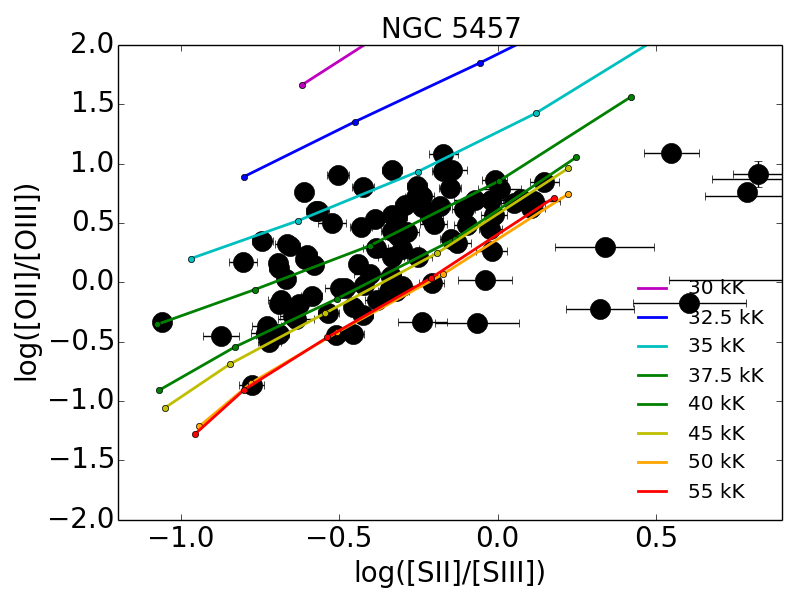}

\caption{Relation between the emission-line ratios [\sii]6717,6731/[\siii]9069,9532 and
[\oii]3727/[\oiii]4959,5007 for the observed galaxies and comparison with models at different
$T_*$ values at the average $Z$ measured in each galaxy.
}

\label{eta}
\end{figure}

\section{Radial variations across galactic discs}

\subsection{Variations of the softness parameter}

We evaluated the radial variations of the derived properties from our
code {\sc HCm-Teff} in the unique available sample in the literature
that provides quality spectroscopic data for a representative 
subsample of \hii\ regions at all galactocentric distances, an accurate 
determination of the chemical abundances from the direct method, and with the
required emission lines necessary for the calculation of the softness parameter
(i.e. [\oii], [\oiii], [\sii], [\siii]).
This is the case of the {\sc chaos} sample, which provides all this information for the
galaxies NGC~628 \citep{chaos1}, NGC~5194 \citep{chaos2}, and NGC~5457 \citep{chaos3}.
We compiled from these works the listed reddening-corrected relative-to-\hb\ appropriate line fluxes and the derived total oxygen chemical abundances as
derived from the direct method. In the case of the [\siii] 9069 \AA\ line we
rescaled its flux using the very close Pa10 line at 9015 \AA\ to \hb\ using the expected theoretical ratio 
at a standard electron density and temperature to minimize reddening and flux calibration uncertainties.
All the compiled \hii\ regions have \hb\ equivalent widths larger than 10 \AA\ in emission,
which ensures they are mostly ionized by very young episodes of star formation \citep{cidfer10}
Overall, we compiled information from 46 \hii\ regions in NGC~628 [44 with a direct determination of 12+log(O/H)],
63 (29) in NGC~5194, and 102 (77) in NGC~5457.

\begin{figure*}
\centering

\includegraphics[width=8cm,clip=]{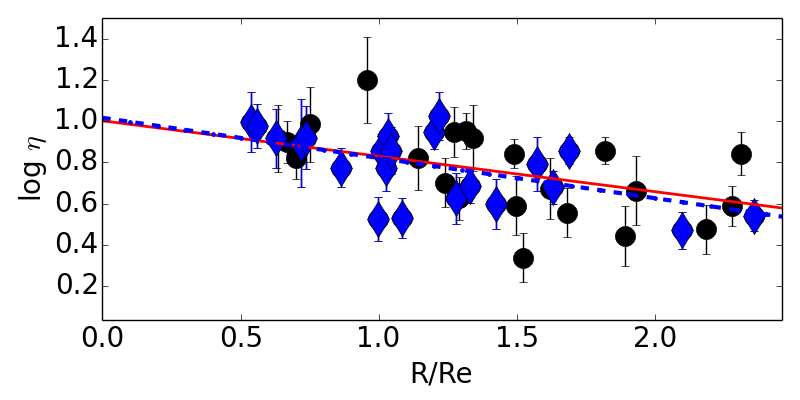}
\includegraphics[width=8cm,clip=]{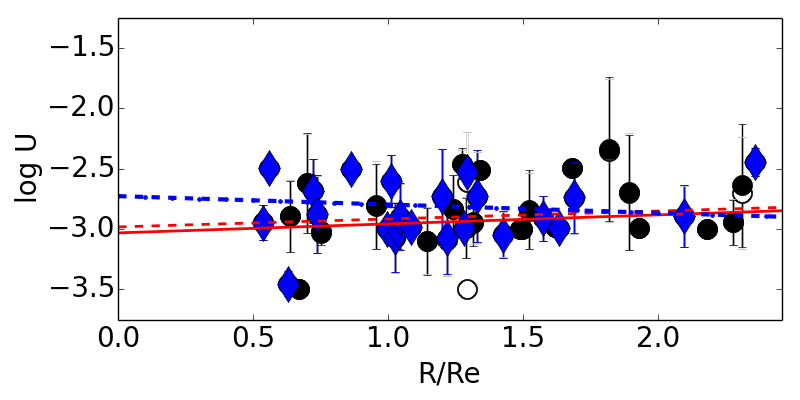}
\includegraphics[width=8cm,clip=]{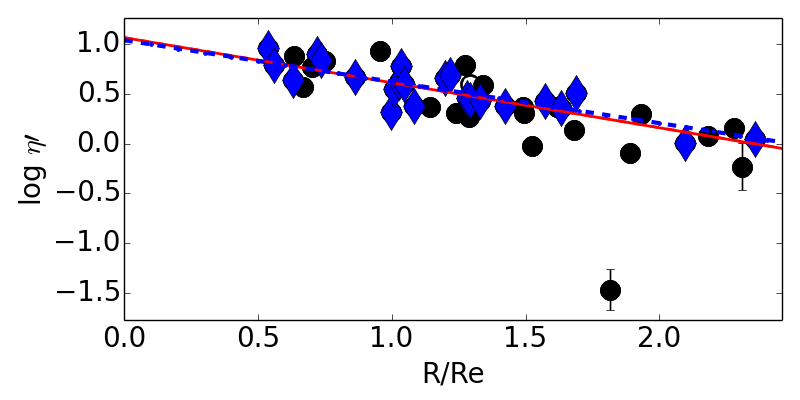}
\includegraphics[width=8cm,clip=]{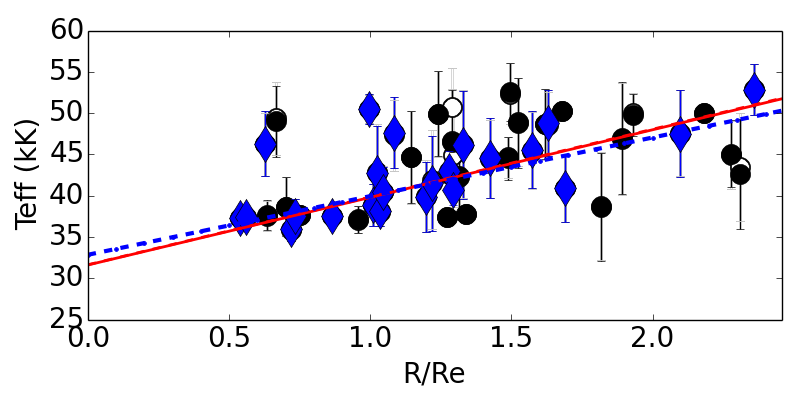}

\caption{Radial distribution and linear fitting normalized to the effective radius for
$\eta$ and $\eta\prime$ (left-hand panels), as derived from observations, and for $T_*$ and log U
(right-hand panels) as derived using our code in NGC 628.
The filled circles represent those \hii\ regions with a direct determination
of the chemical abundance and its corresponding linear fitting is plotted with a dashed line. The white circles represent all the other regions whose O/H
was derived using other methods. The linear fitting to the whole sample is represented with a solid line. Finally, blue diamonds represent those regions whose H$\alpha$ luminosity is
larger than the average. The linear fitting to this subsample is represented with a blue dotted-dashed line.}

\label{n628_grad}
\end{figure*}

\begin{figure*}
\centering

\includegraphics[width=8cm,clip=]{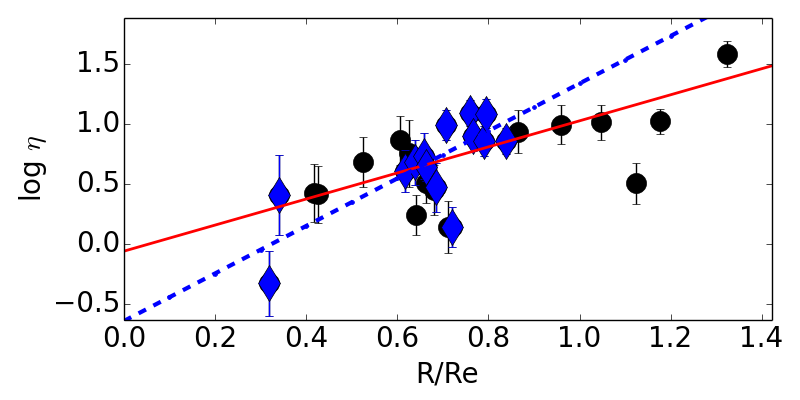}
\includegraphics[width=8cm,clip=]{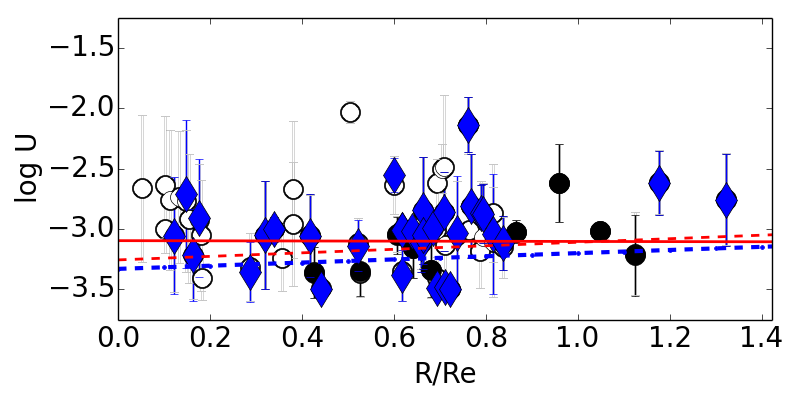}
\includegraphics[width=8cm,clip=]{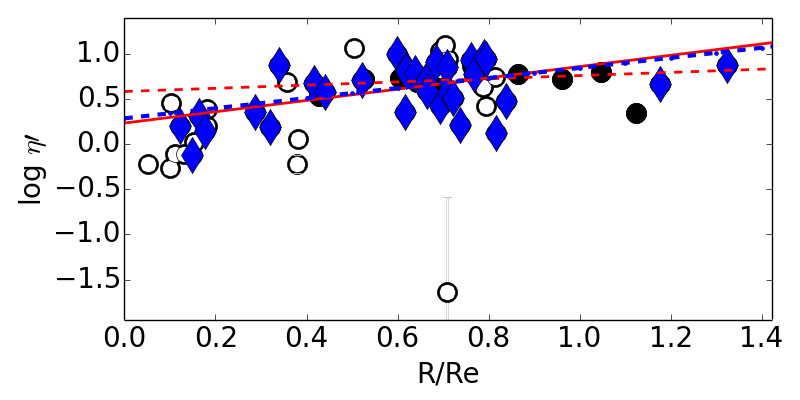}
\includegraphics[width=8cm,clip=]{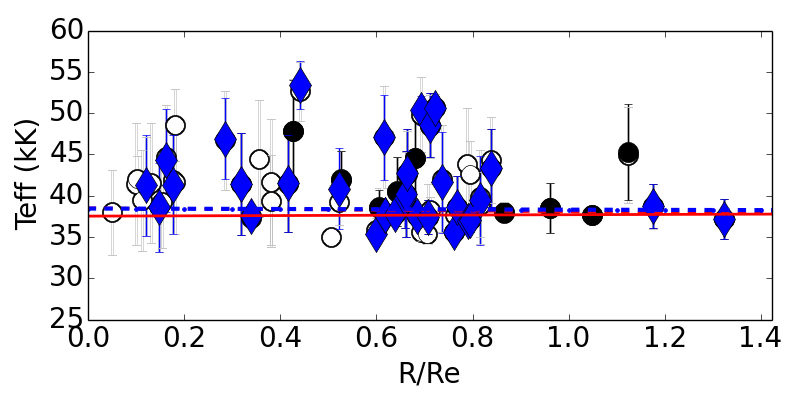}

\caption{Same plots as those described in Fig.\ref{n628_grad} for the galaxy NGC 5194.}

\label{n5194_grad}
\end{figure*}

\begin{figure*}
\centering

\includegraphics[width=8cm,clip=]{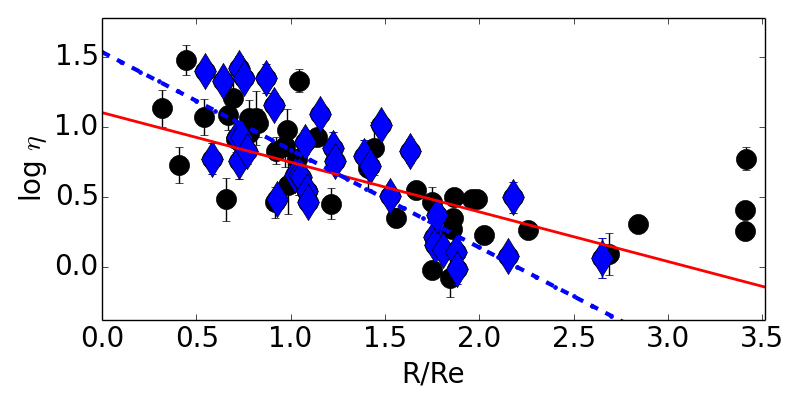}
\includegraphics[width=8cm,clip=]{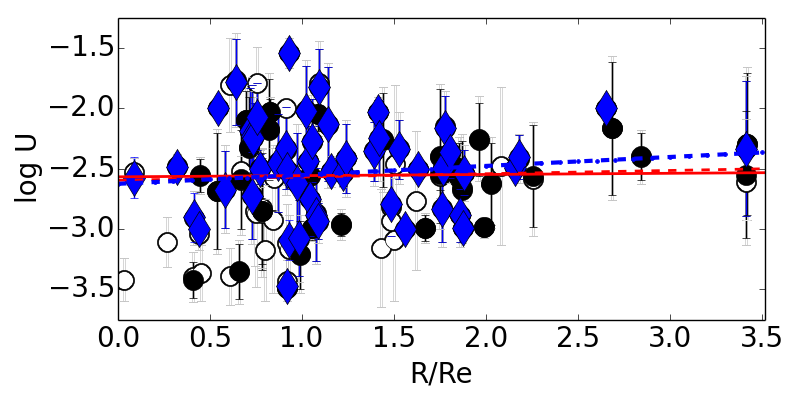}
\includegraphics[width=8cm,clip=]{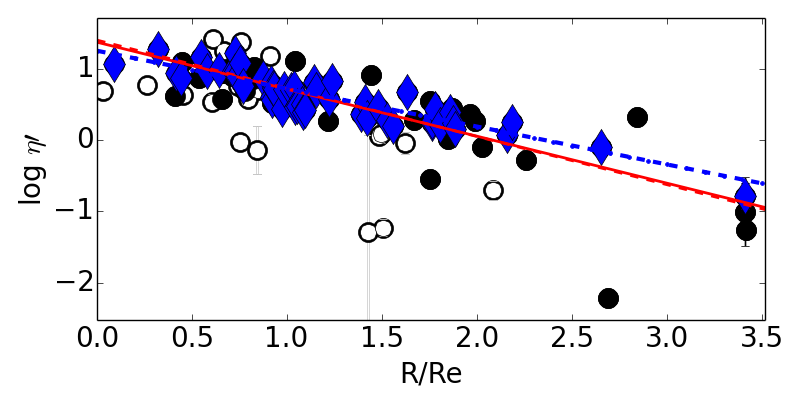}
\includegraphics[width=8cm,clip=]{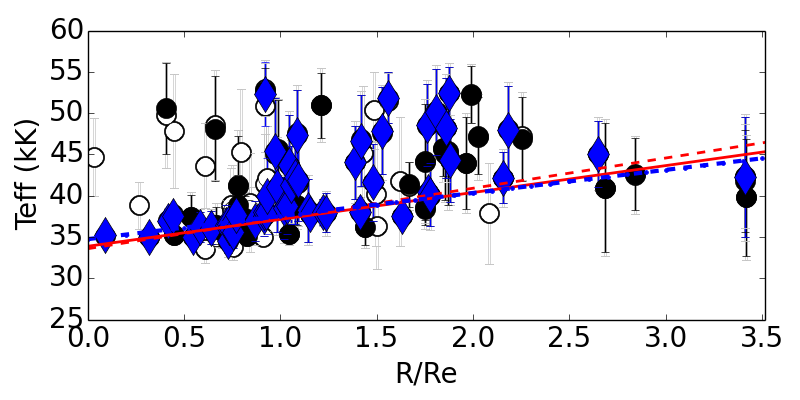}

\caption{Same plots as those described in Fig.\ref{n628_grad} for galaxy NGC 5457.}

\label{n5457_grad}
\end{figure*}

\begin{figure}
\centering

\includegraphics[width=8cm,clip=]{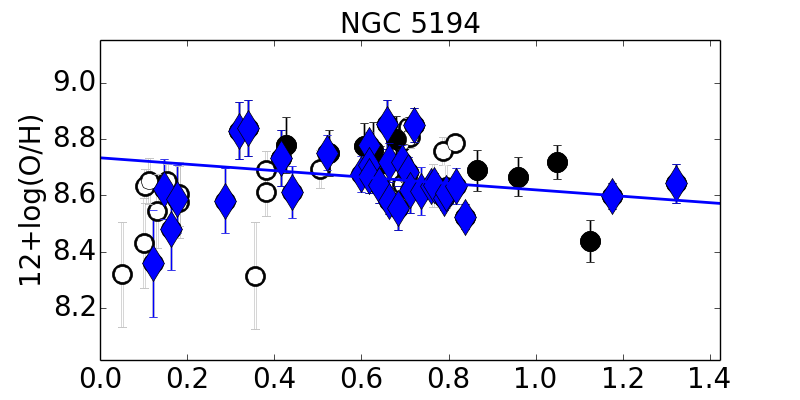}
\includegraphics[width=8cm,clip=]{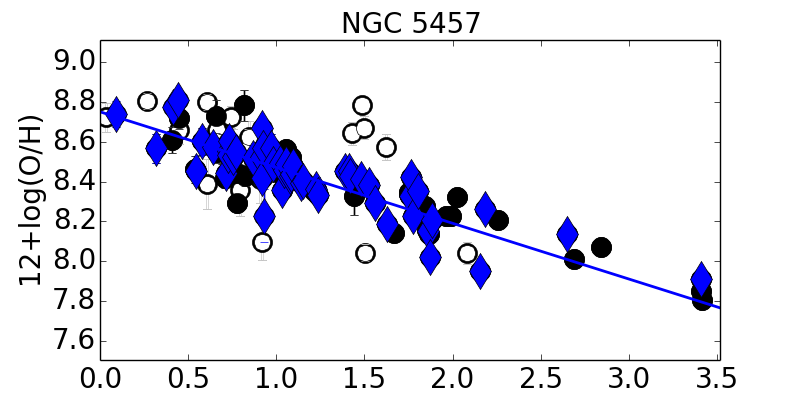}
\includegraphics[width=8cm,clip=]{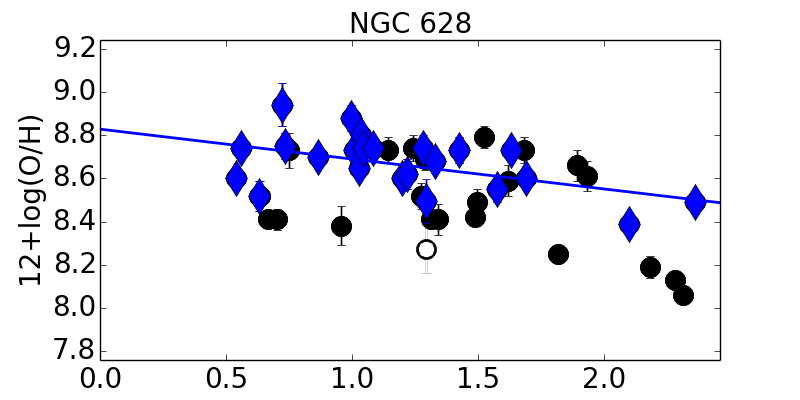}

\caption{Radial distribution of the total oxygen abundances in the \hii\ regions of the three {\sc chaos} galaxies. We show
NGC~628 in the top panel, NGC~5194 in the middle panel, and NGC~5457 in the bottom panel. The radial scale is normalized to the effective radius in each object.
The black circles represent regions whose abundances were derived using the direct method, the white circles have abundances derived using the {\em HCM} code, while diamonds 
represent regions with an \ha\ luminosity larger than the average in each galaxy. The red solid line represents the error-weighted linear fitting to the 
whole sample. The dashed line represents the fitting to the regions with abundances derived 
following the DM, and the dot-dashed lines the fitting to the most luminous regions.}

\label{grad_O}
\end{figure}

All these nearby star-forming disc galaxies present a spatial variation of $Z$ 
and, at the same time, different values of the hardening of the ionizing field of radiation
in different positions of the galaxies as already described in \cite{pmv09}. In Figure \ref{eta}
we show the position of the studied \hii\ regions of the three galaxies in the plane
log([\sii]/[\siii]) vs log([\oii]/[\oiii]). As reference we also plotted curves of equal $T_*$
from the models described above at the  average $Z$ in each galaxy.

We can see that there is a good  agreement between the position of
the observations in the three galaxies and the space covered by the grid of models.
Nevertheless there is a slight mismatch of some \hii\ regions lying below the space covered 
by the models that exceeds the maximum considered $T_*$. This partial disagreement 
between models and observations was already pointed out in \cite{pmv09}, notably for star-forming dwarf galaxies,
and could be due to problems in the atomic coefficients used by the models, the stellar libraries,
or to an oversimplification in the geometry of the models as the matter-bounded case is not assumed in the grid
and this can largely affect the relative intensities of consecutive ionic stages across
the radius of the ionized gas.

We explored the radial variation of the softness parameter in the three galaxies
taking the corresponding deprojected galactocentric distance of each \hii\ region. We analysed both the
$\eta$ parameter, based on ionic abundances only for those \hii\ regions with a direct estimation of the
chemical abundance, and the $\eta\prime$ parameter, using only emission lines in all \hii\ regions.
The results, normalized to the effective radius, are shown in Figure \ref{n628_grad} for NGC~628, Figure \ref{n5194_grad} for NGC~5194, and
Figure \ref{n5457_grad} for NGC~5457. 

In Table \ref{slopes} we list the slopes of the corresponding linear fittings performed
to the radial distribution of these parameters considering the weights of the
associated errors as propagated from the reported 
relative intensities and ionic abundances.
The fittings were calculated using the radii normalized to the effective radius in each object.

Both NGC~628 and NGC~5457 show very clear decreasing trends of $\eta^\prime$, while NGC~5194 presents
an increasing radial variation, although sensibly flatter when we consider only those \hii\ regions with a
direct determination of the oxygen abundance. These trends are confirmed when we explore the radial variations of $\eta$,
which is possible in most of the \hii\ regions of the {\sc chaos} sample due to the precision in the determination of the involved ionic
abundances in each \hii\ region. The radial variations in these three objects
were already observed by \cite{pmv09} using a lower number of \hii\ regions.

We also checked whether these radial trends are affected by the luminosity of the regions, as $T_*$,
one of the quantities governing the behavior of the softness parameter, could be affected by the statistical fluctuations in
low-scale regions (\citealt{cervino2000}). We considered the mean H$\alpha$ luminosity of the \hii\ region distribution in
each galaxy as a lower limit to look for differences with respect to the total sample. We found the values 10$^{39.8}$ erg/s in NGC~628,
10$^{40.1}$ erg/s in NGC~5194, and 10$^{40.1}$ erg/s in NGC~5457.
Taking {\sc starburst99} \citep{sb99} models as a reference, these luminosities would correspond to cluster masses of about
3$\times$10$^3$  M$_{\odot}$ in NGC~628 and 6$\times$10$^3$ M$_{\odot}$ in NGC~5194 and NGC~5457, assuming an instantaneous burst 
at $Z$ = $Z_{\odot}$ and standard conditions of the cluster.
As can be seen in the corresponding figures and in Table \ref{slopes} the observed trends 
do not vary significantly even when we only consider the top half of the H$\alpha$ distribution of the \hii\ regions in these galaxies.

Nevertheless we still have to explore whether these radial variations of both softness parameters can
be associated with a real radial variation of the hardening of the 
ionizing stellar radiation field across the galactic discs.

\begin{table*}
\centering
\caption{Slopes and errors of the linear fittings for the studied galaxies
normalized by the corresponding effective radius. We consider three different subsamples of regions in
each galaxy: (i) the whole sample, (ii) only \hii\ regions with a direct determination of the chemical abundance, and (iii) regions
with H$\alpha$ luminosity larger than the average in each galaxy.}

\begin{tabular}{lccccccc}
\hline
\hline
Object & Selection & Regions & $\alpha$ (log $\eta$) & $\alpha$ (log $\eta\prime$) & $\alpha$ (log U) & $\alpha$ (T$_*$) & $\alpha$ (12+log(O/H))\\
         &  (DM)    &    & (dex/R$_e$)    &  (dex/R$_e$)   &  (dex/R$_e)$  & (kK/R$_e$)  & (dex/R$_e$) \\
\hline
NGC~628  & {\em All} & 46 &   -- & -0.45 $\pm$ 0.05 & 0.07 $\pm$ 0.10 & 8.2 $\pm$ 0.7 & -0.21 $\pm$ 0.05 \\
       & {\em O/H from DM}  & 44 &   -0.17 $\pm$ 0.05  & -0.45 $\pm$ 0.05 & 0.07 $\pm$ 0.09 & 8.2 $\pm$ 0.7  & -0.21 $\pm$ 0.05\\
       & {\em L(H$\alpha$) $>$mean}  & 23 & -0.19 $\pm$ 0.06  & -10.41 $\pm$ 0.07 & -0.07 $\pm$ 0.15 & 7.1 $\pm$ 1.8  & -0.14 $\pm$ 0.04 \\
\hline
NGC~5194 & {\em All} & 68 & -- & 0.62 $\pm$ 0.11 & -0.01 $\pm$ 0.19 & 0.2 $\pm$ 1.8 & 0.01 $\pm$ 0.06 \\
       & {\em O/H from DM}  & 29 &  1.08 $\pm$ 0.21 &  0.18 $\pm$ 0.16   & 0.15 $\pm$ 0.28 &-0.5 $\pm$ 2.6  & -0.21 $\pm$ 0.08\\
       & {\em L(H$\alpha$) $>$mean}  & 32 &  1.98 $\pm$ 0.63 & 0.55$\pm$ 0.15 & 0.13 $\pm$ 0.38 &  -0.1 $\pm$ 4.7 & -0.11 $\pm$ 0.07  \\
\hline
NGC~5457  & {\em All} & 102 &  --  & -0.66 $\pm$ 0.05 &  0.01 $\pm$ 0.07 & 3.3 $\pm$ 0.5  & -0.27 $\pm$ 0.02 \\
       & {\em O/H from DM}  & 78 &-0.35 $\pm$ 0.05  & -0.68 $\pm$ 0.05  & 0.03 $\pm$ 0.08 & 3.6 $\pm$ 0.6  & -0.26 $\pm$ 0.02 \\
  & {\em L(H$\alpha$) $>$mean}  & 52 & -0.70 $\pm$ 0.08 & -0.53 $\pm$ 0.04 & 0.07 $\pm$ 0.09 & 2.8 $\pm$ 0.6 & -0.28 $\pm$ 0.02  \\
\hline

\label{slopes}
\end{tabular}
\end{table*}

\subsection{Radial variation of $U$ and $T_*$}

In order to check whether the obtained radial variations of the 
softness parameter in the three studied galaxies can be really interpreted as
a radial variation of the hardening of the ionizing stellar field of radiation, we 
applied {\sc HCm-Teff}, described in Section 2, to
the sample of \hii\ regions.

We considered the oxygen abundances derived following the direct method given by {\sc CHAOS} 
to constrain the grid of models in those \hii\ regions where it was possible.
For the rest of the \hii\ regions without a direct estimation of $Z$ we resorted to the code {\sc Hii-Chi-mistry} v.3.0
({\sc HCm}; \citealt{hcm}) to provide an estimation of 12+log(O/H) for
the calculation of both $U$ and $T_*$.
This code follows a very similar methodology to the one described in this work as it is based
on a Bayesian-like approach to establish a comparison between the measured optical emission lines to
the values predicted by a large grid of photoionization models.
The resulting abundances  are consistent with the direct method even in the absence of any auroral line although
the uncertainty in this case is larger.
In addition the error fluxes of the used emission lines are taken into account by the code
for the final error calculation following a Monte Carlo iteration.
We checked that the oxygen total abundances derived using the {\sc HCm} code in the absence of the [\oiii] 4363 \AA\ line agree with
the abundances derived following the direct method in those {\sc chaos} \hii\ regions with any auroral line with a dispersion lower than 0.2 dex and without any systematic offset even in the regime of oversolar metallicity.
Although different authors point towards a systematic offset of around 0.2 dex between the abundances derived from the
direct method and those coming from photoionization models (i.e. \citealt{blanc15, asari16}), the {\sc Hii-Chi-mistry} code
obtains 
no discrepancies by means of an assumed empirical relation between 
metallicity and excitation. The agreement with the \hii\ regions of the {\sc chaos} sample is also discussed
in \citealt{pm16}.

We calculated the slopes of the error-weighted linear fittings of log $\eta\prime$,
12+log(O/H), log $U$ and $T_*$ and we normalized to the effective radius (R$_e$) (i.e. the radius encompassing half of the total luminosity).
The slopes of the linear fittings and
their corresponding errors are listed in Table \ref{slopes}.

For the sake of consistency, we provide the results of the radial variations and fittings both for $U$ and  $T_*$ and  for those
\hii\ regions only with a direct determination of O/H to explore whether the obtained dispersions can be caused due to
the factor of including objects with abundances obtained from other methods.
We also show in Figure \ref{grad_O} the radial profiles of the total oxygen abundance in the three
galaxies, taking the whole analysed sample and only for regions with a direct estimation of the electron temperature. 
Clear negative radial variations of the metallicity are observed in the three objects, with slopes larger than -0.2 dex/R$_e$,
in agreement with the results obtained by the {\sc chaos} collaboration. Only in the case of NGC~5194 do we obtain a much flatter slope when
the \hii\ regions in the central regions of this galaxy are considered using the values derived by {\em HCM}.

For the calculation of $T_*$ in those \hii\ regions that lie in the plane
[\sii]/[\siii] vs. [\oii]/[\oiii] below the space covered by the grid of models,
we checked that the code always assigns to them 
values larger than 50k K, but we cannot confirm in
these cases whether the position of these regions in the plot can
be due to other factors (i.e. different geometry, photon leakage).

The distributions of the derived log $U$ and $T_*$ for \hii\ regions in the
three galaxies across the deprojected galactocentric distances normalized to the effective
radius are plotted in Figure \ref{n628_grad} for NGC~628, Figure \ref{n5194_grad} for NGC~5491, 
and Figure \ref{n5457_grad} for NGC~5457. 

Though the radial behavior of $\eta$ is different in the three galaxies, they show a flat radial 
slope of the ionization parameter so the \hii\ regions studied here present a similar 
ionization parameter, on average, independently of their galactocentric distance. 
We do not observe any great difference between the slopes fitted to all regions or just to
the regions with a direct determination of the abundance or to the most luminous ones. The slope only 
appears to be slightly higher in NGC~5194 when only regions with a direct determination of O/H are considered.
At the same time the \hii\ regions in the three galaxies present average excitation properties that are in 
agreement with their sizes and average $Z$, taking the characteristic value of the obtained linear fitting 
at the effective radius. In this way NGC~628 has a characteristic log $U$ value of -2.96,
NGC~5194 has -3.11, and NGC~5457 has -2.56.

Regarding $T_*$ we observe radial
variations consistent with those observed in the softness parameter, although 
the sign of the slope observed in $\eta$ cannot be always interpreted as radial variations of $T_*$
in a determined sense.
On one hand the two galaxies with clear decreasing slopes of the $\eta$ parameter, NGC~628 and NGC~5457,
present at the same time very clear increasing slopes of $T_*$. 
Therefore in these two galaxies the observed variation of the {\em softness} parameter can be
directly associated with a hardening of the stellar radiation across the galactic disc.
On the other hand, NGC~5194, presenting increasing slopes of $\eta$, has in average a flat radial distribution of the
$T_*$ values.
These flat radial distributions cannot be related to the large number of \hii\ regions not covered 
by the grid of models in this galaxy. Considering only the 35 \hii\ regions with values of [\oii]/[\oiii] and [\sii]/[\siii]
values reproduced by our grid, the slope of the linear fitting for $T_*$ is 500 $\pm$ 600 K/R$_e$, and for log $U$ it is -0.22 $\pm$ 0.26 dex/R$_e$.

Neither the obtained slopes nor the associated dispersions vary significantly when only \hii\ regions with
a direct determination of the abundance or regions belonging to the top half of the H$\alpha$ luminosity distribution are considered
in the three galaxies.

Therefore, though the direct analysis  of the radial variation of the softness parameter in these three 
well-studied disc galaxies cannot be
directly interpreted as equivalent radial variations of $U$ or $T_*$  we find some correlation between the obtained slopes. 
Moreover the results evidence the need for a detailed analysis of all involved 
quantities in the softness parameter
instead of using this parameter as a direct indicator of the properties of an \hii\ region.

\subsection{Correlation with other properties}

\begin{table*}
\centering

\caption{Slopes of the linear fittings to log $\eta\prime$ and 12+log(O/H) as measured, and log $U$ and $T_*$ as derived using our code 
as a function of the galactocentric distance in units of kpc for the {\sc chaos} galaxies studied in this work and
for the sample analyzed in \refpmv. We provide the references
of the observed, used emission lines in each object.}

\begin{tabular}{lccccc}

\hline
\hline
Object & $\alpha$ (log $\eta\prime$) & $\alpha$ (O/H)   & $\alpha$ (log $U$) & $\alpha$ ($T_*$) & Ref. \\
  & dex/kpc &  dex/kpc &dex/kpc & K/kpc   & \\
\hline
NGC~300 & -0.059 $\pm$ 0.030 & -0.008 $\pm$ 0.006 & 0.035 $\pm$ 0.029 & 1900 $\pm$ 800 &  \cite{bresolin09} \\
NGC~598 & -0.175 $\pm$ 0.023 & -0.016 $\pm$ 0.004 & 0.105 $\pm$ 0.039 & 1100 $\pm$ 300 & \cite{vilchez88} \\
NGC~628& -0.062 $\pm$ 0.015 & -0.029 $\pm$ 0.010 & 0.017 $\pm$ 0.023 & 1800 $\pm$ 200 & \cite{chaos1} \\
NGC~1232 &  -0.010 $\pm$ 0.018 & -0.002 $\pm$ 0.029 & 0.052 $\pm$ 0.008 & -100 $\pm$ 200 &  \cite{bresolin05} \\
NGC~1365  & -0.010 $\pm$0.005 & 0.002 $\pm$ 0.005 & -0.003 $\pm$ 0.007 & -100 $\pm$ 100 & \cite{bresolin05}  \\
NGC~2403 & -0.083 $\pm$0.022 & -0.051 $\pm$ 0.028 & -0.039 $\pm$ 0.031 & 2100 $\pm$ 300 & \cite{garnett97}  \\
NGC~2903  & -0.003 $\pm$ 0.002 & -- & 0.047 $\pm$ 0.040   & 500 $\pm$ 600   & \cite{bresolin05}\\
NGC~2997  & -0.002 $\pm$ 0.001 &  0.000 $\pm$ 0.001 & 0.030 $\pm$ 0.055   & 800 $\pm$ 600    &  \cite{bresolin05} \\
NGC~5194 & -0.118 $\pm$ 0.116 & -0.042 $\pm$ 0.025 & 0.063 $\pm$ 0.055 & -300 $\pm$ 100 & \cite{bresolin04}\\
NGC~5194 & 0.078 $\pm$ 0.012 & 0.003 $\pm$ 0.006 &0.001 $\pm$ 0.002 & 0 $\pm$ 200 & \cite{chaos2}\\
NGC~5236 & -0.084 $\pm$ 0.046 & -0.052 $\pm$ 0.028 & 0.012 $\pm$ 0.052 & 400 $\pm$ 500 & \cite{bresolin05}    \\
NGC~5457 & -0.032 $\pm$ 0.009 & -0.019 $\pm$ 0.005 & 0.002 $\pm$ 0.003 & 400 $\pm$ 100   &   \cite{kennicutt03}\\
NGC~5457 & -0.077 $\pm$ 0.005 & -0.030 $\pm$ 0.002 & 0.001 $\pm$ 0.007 & 400 $\pm$ 100 & \cite{chaos3}  \\
\hline

\end{tabular}
\label{table2}
\end{table*}

\begin{figure*}
\centering

\includegraphics[width=8cm,clip=]{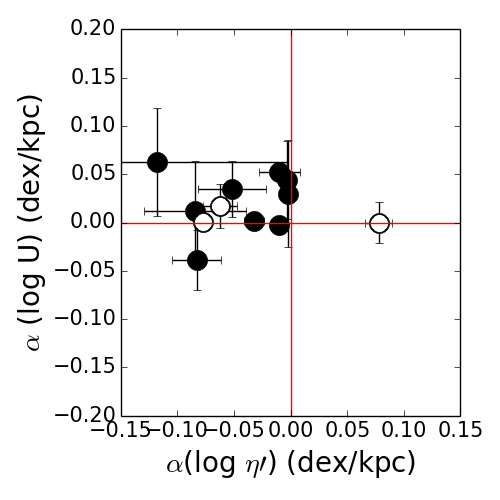}
\includegraphics[width=8cm,clip=]{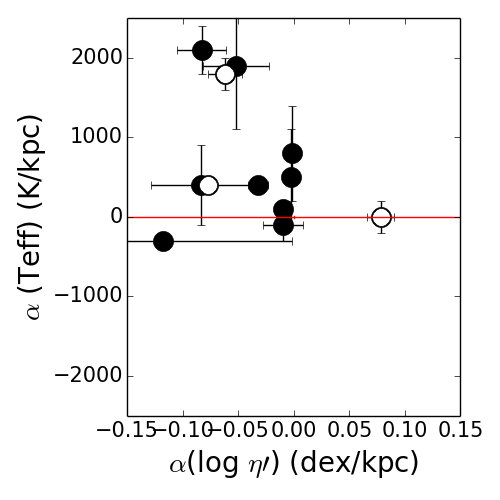}
\includegraphics[width=8cm,clip=]{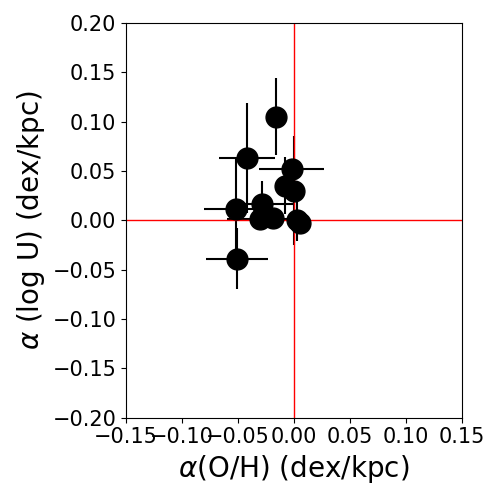}
\includegraphics[width=8cm,clip=]{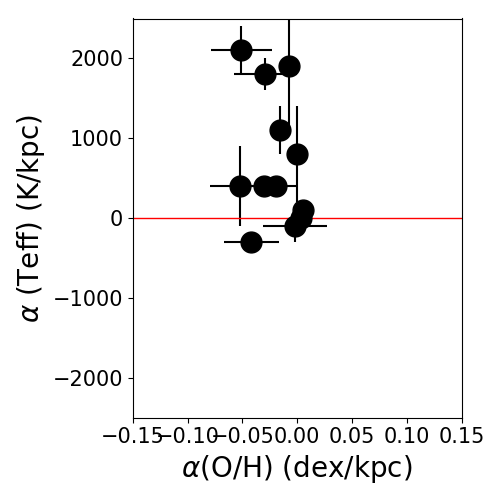}

\caption{Relation between the slopes of the linear fittings and the radial distributions of several
properties in the studied galaxies. 
The upper row represents the relation between the slopes of log $\eta\prime$ 
and log $U$ and T$_*$ while in  the lower row we show  the slopes of 12+log(O/H) with
the same derived properties.
The white circles represent the {\sc chaos} sample and the
black circles the sample studied in \refpmv.}

\label{comp_slopes}
\end{figure*}

Once the different dependences of the softness parameter on the $U$
and $T_*$ have been analysed in the three studied objects, 
it is possible to re-examine the radial variations
of these properties as a function of the slopes of other well-known derived quantities in the same galaxies and to 
establish correlations with other integrated properties.

In any case, the three-galaxy sample with good-quality data from the {\sc chaos} collaboration is not enough to
extract general conclusions about the behavior
of these radial variations. In fact, only two out of the three galaxies present the decreasing gradient 
of the softness parameter already observed in the majority of the sample studied in \cite{pmv09}.
Of the nine analysed galaxies with available data to study the radial variation of
the softness parameter in \cite{pmv09}, most present a clear decreasing radial variation.
For that reason, although that sample has a much lower number of
\hii\ regions in each object and the determination of chemical abundances in each
object is more uncertain, we reanalysed this sample in order to
inspect the causes of the radial variation of the softness parameter.

In Table \ref{table2} we give the results of the application of the {\sc HCm-Teff} code
to those \hii\ regions presenting the lines involved in the calculation of the
softness parameter. We also list the three CHAOS objects studied above
multiplying the obtained slopes by the effective radius
to give them in dex kpc$^{-1}$ as in the rest of the objects.
We also list in the same table the slopes of the linear fittings in the same \hii\ regions 
corresponding to the radial distribution of log $\eta\prime$ and 12+log(O/H), as derived from the direct method
in those \hii\ regions with at least an auroral emission line.

As can be seen, all of them 
present a positive or flat gradient of log $U$ with the exception of NGC~1365,
which presents a value compatible with a flat slope, and NGC~2403, in this 
case with a large dispersion in the obtained slope.
In the same way all galaxies present a clear positive slope in the gradient of $T_*$ or
values compatible with a flat slope, but in none of them do we obtain clear
evidence of a negative slope of $T_*$ across the
galactic discs.

In Figure \ref{comp_slopes} we show a comparison between the slopes of the softness parameter,
as calculated using log $\eta\prime$, and the slopes for both $U$ and $T_*$
to inspect which physical property has a larger weight.
Although it is not possible 
to extract a clear correlation because of the shortage of
the analysed sample, there is a large coincidence between
the sign of the slope of $\eta\prime$ and log $U$ and $T_*$
for almost all the objects.
The general picture is that of a galaxy with a positive gradient of the softness parameter, which
translates from the models in a positive or flat gradient for $U$ and $T_*$ but
with no clear evidence about which of them has a larger weight on the softness parameter.

Most of the disc galaxies present at the same time lower O/H values at larger galactocentric distances 
as studied by the authors presenting the spectroscopic studies of the corresponding \hii\ regions.
This could be related to the existence of an excitation variation with distance
or with a lower value of $T_*$ in the innermost regions, where the amount of metals is higher.
To evaluate the existence of these correlations, we show in the lower
panels of Figure \ref{comp_slopes} the relation between the slopes of the linear fittings 
(as a function of radius in kpc) of the total oxygen abundance and log $U$ and $T_*$. 
As in the case of $\eta$ there is a
general concordance between the sign of the O/H slopes and those of excitation 
and hardening of the radiation but, on the other hand, it is not possible to extract the main
source of correlation.

\begin{figure*}
\centering
\includegraphics[width=8cm,clip=]{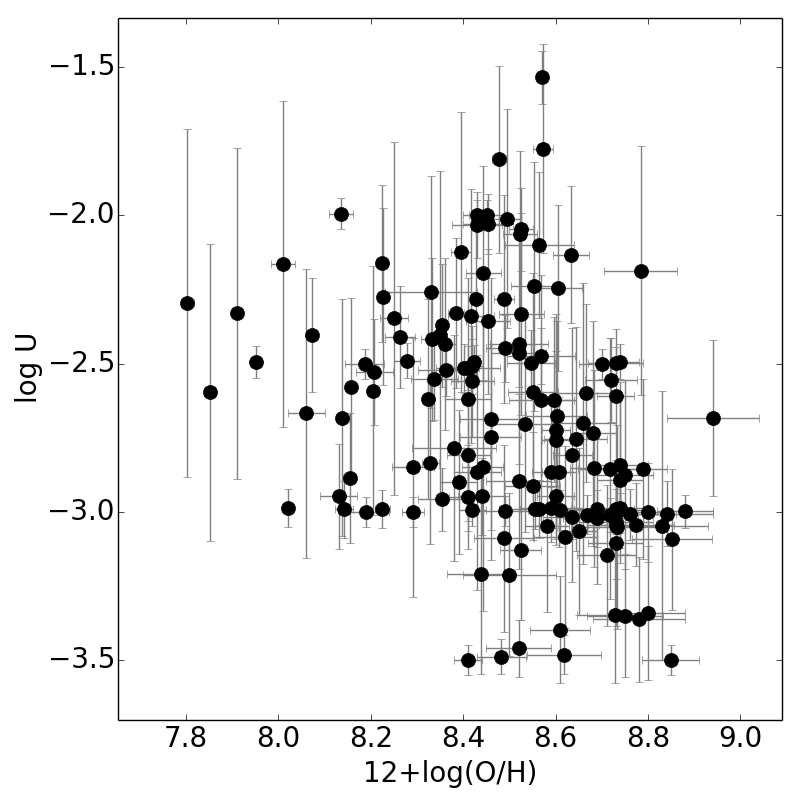}
\includegraphics[width=8cm,clip=]{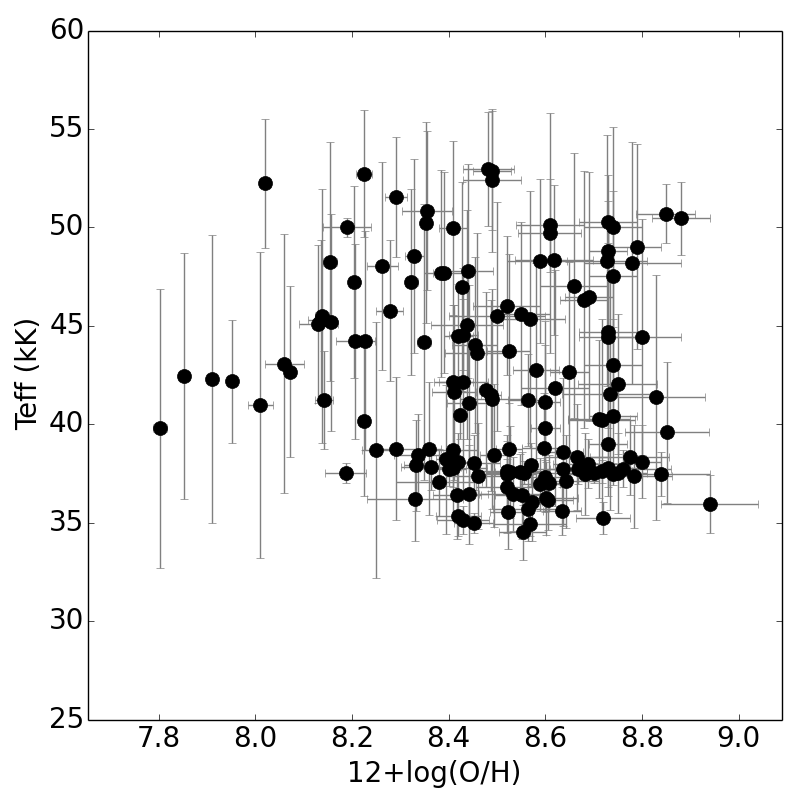}
\caption{Relation between the total oxygen abundance as derived by the direct method and $U$ (left-hand panel) and
$T_*$ (right-hand panel) derived from {\sc HCm-Teff} in those \hii\ regions studied in the three {\sc chaos} galaxies
with the appropriate emission lines.}
\label{rel-oh}
\end{figure*}

The existence of a correlation between $Z$ and $U$,
in the sense that regions with a larger metal content are on average less excited, 
has been widely described and discussed in the literature (e.g. \citealt{hcm}) 
but it has been interpreted in terms of the additional dependence  
on $T_*$ of the emission-line ratios used to derive $U$. 
In fact, when we supply a methodology to resolve the
dependence of emission lines on both $U$ and $T_*$ we appreciate a certain correlation 
between $Z$ and both the gas excitation and the hardness of the radiation. 
In Figure \ref{rel-oh} we show the relation between the total oxygen 
abundance as derived by the direct method in the \hii\ regions of the
three {\sc chaos} galaxies and $U$ and $T_*$ as derived using the {\sc HCm-Teff} code.
 We limit this plot to this sample as these \hii\ regions present a very accurate abundance determination from one or several
auroral emission lines and were observed under much more homogeneous conditions.
No clear general correlations are shown by the data in either plot; nonetheless, the influence of 
other HII region parameters (e.g. evolution of the ionizing clusters; geometry, multiple 
versus compact ionizing sources, luminosity/size) could be operating for each sample.

\begin{figure*}
\centering

\includegraphics[width=8cm,clip=]{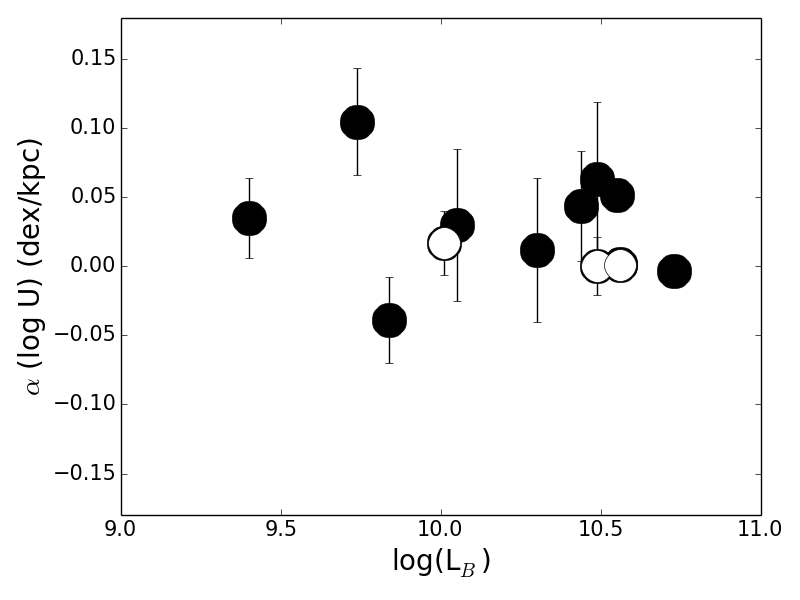}
\includegraphics[width=8cm,clip=]{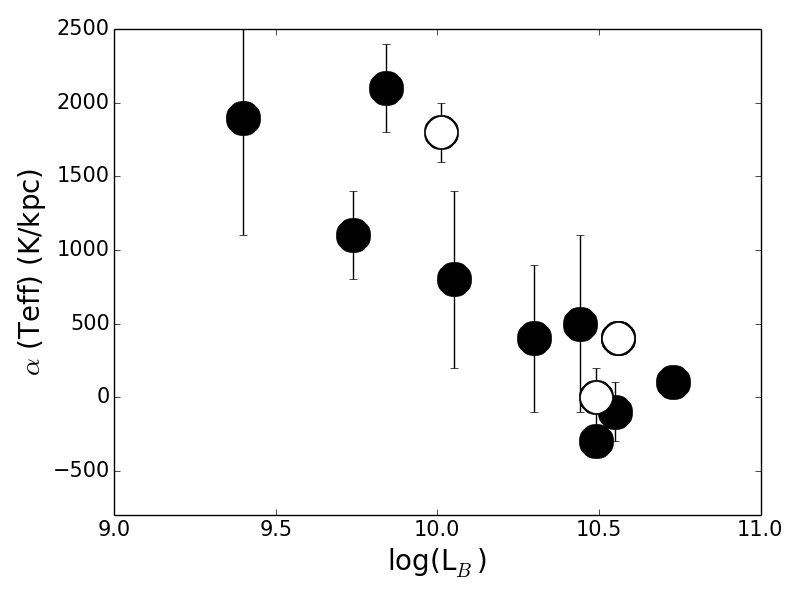}

\caption{Relation between integrated luminosity in the $B$ band and the slopes of
the linear fittings to log $U$ (left) and $T_*$ (right). The white circles correspond to the three {\sc chaos} galaxies, 
while the black circles represent the results from the galaxies presented in \refpmv.}

\label{MLB}
\end{figure*}

\begin{figure*}
\centering

\includegraphics[width=8cm,clip=]{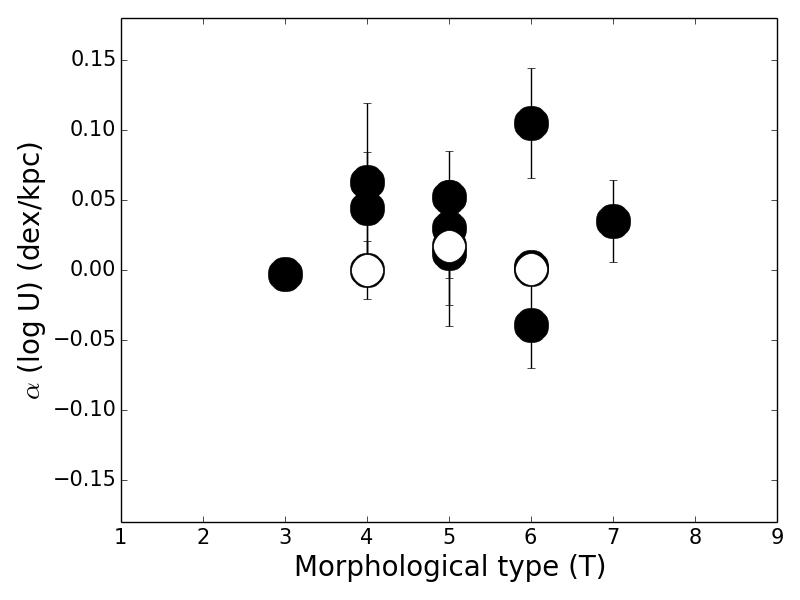}
\includegraphics[width=8cm,clip=]{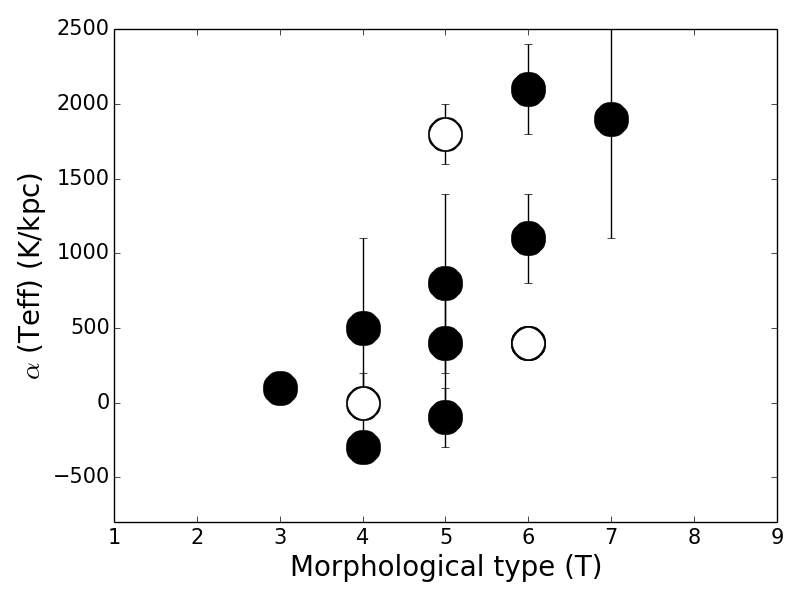}

\caption{Relation between the Hubble morphological type  and the slopes of
the linear fittings to log $U$ (left) and $T_*$ (right). The white circles correspond to the three {\sc chaos} galaxies, while the black circles
represent the results from the galaxies presented in \refpmv.}

\label{morph}
\end{figure*}

Another interesting test to check the existence of a connection between 
excitation, radiation hardness,  and $Z$ is exploration of  
the relation between  the average radial variation of the 
softness parameter and the integrated luminosity of the galaxies.
 \cite{pmv09} pointed out a certain correlation between the luminosity of galaxies and the slope of the linear fittings
of log $\eta\prime$, as it has been observed in the case of $Z$.
With the current study it is possible to figure out whether this correlation is mostly due to the dependence
of the softness parameter on $U$ or on $T_*$.
In Figure \ref{MLB} we show this relation both for $U$ and for $T_*$ though no global trend can be
extracted owing to the scarcity of our sample. 
Despite the large dispersion, it is possible to extract some average trends. In this way for those galaxies with log $L_B$ $<$ 10.2 erg/s 
the average log $U$ slope is 0.03 ($\pm$0.04) dex/kpc, while
for galaxies with log $L_B$ $>$ 10.2 erg/s it is 0.02 ($\pm$ 0.02) dex/kpc, both values consistent to within the errors.
However, a much clearer correlation is found when we perform the same analysis regarding $T_*$ as the slope for less 
bright galaxies (log $L_B$ $<$ 10.2 erg/s) is 1500 $\pm$ 400 K/kpc, while for the brighter galaxies
(log $L_B$ $>$ 10.2 erg/s) is the mean value is only 400 $\pm$ 300K/kpc. This difference in the observed slopes appears well 
established; thus in order to derive a precise value a study with a larger number of galaxies will shed more light on
the gradient of the hardening of the field of radiation in disc galaxies.

In the same way we explored the possible existence of a correlation between the
slopes of these gradients and the Hubble morphological type. In \cite{pmv09} it appeared that galaxies with a later
type present on average more pronounced $\eta\prime$ slopes. In Figure \ref{morph} we now show this relation once the dependence of $\eta\prime$ has been divided in
$U$ and $T_*$.
As in the case of the total integrated luminosity the existence of a possible correlation appears now only in the case
of the radial variation of the harness of the ionizing radiation. Galaxies with later Hubble types (T $>$ 5) have
larger $\alpha_{T*}$ (1100 $\pm$ 500 K/kpc) than those with
earlier types (T $\leq$ 5) which have on average 400 $\pm$ 500 K/kpc.
On the contrary no noticeable difference is found for the case of the ionization parameter.
This difference in the average slopes of $T_*$ as a function of the morphological type can just be the
effect of a bias in the luminosities of the galaxies for different types. Indeed the mean total luminosity in the $B$ band for late types
is lower (log $L_B$ = 10.0 erg/s) than for the early ones (10.4 erg/s).

Several works have found a clear correlation of stellar mass, stellar mass surface density, and luminosity with morphological types \citep{GonzalezDelgado:2015,Garcia-Benito:2017}, in the sense that early-type galaxies on average are more massive, redder, and show larger absolute magnitudes.

\section{Summary and conclusions}

In this work we have presented a new methodology to derive $U$ and $T_*$
in \hii\ regions. These two parameters are key to understand many
of the observed relations between the relative fluxes of their most prominent emission lines.

Our approach departs from the relation between the so-called {\em softness} parameter with both $U$ and $T_*$. 
In this case we explore the two-dimensional space of the plane [\sii]/[\siii]
versus [\oii]/[\oiii] emission-line ratios in the optical and near-infrared spectral range to figure out
the combination of them that leads to the observed patterns in $\eta$. According to photoionization models there is a non-negligible dependence on
$Z$ in this plane that must also be resolved.
We used a Bayesian-like code called {\sc Hcm-Teff} that compares the prediction of
a large grid of photoionization models in this plane with the observed emission-line ratios. The code also allows us to use only
a given pair of lines but as the slope of the curves with the same $U$ and $T_*$ is not 1,
[\oii]/[\oiii] leads only to good estimations of $T_*$ while [\sii]/[\siii] is
more appropriate for log $U$.

Among the limitations that the used grid of models can present when its results are
compared to observations we can identify the use of a central single-star ionizing 
spectral distribution, the spherical matter-bounded geometry of the surrounding gas,  
and the constancy of density or dust-to-gas ratios. A more sophisticated exploration of
these quantities in the models can lead to different degeneracies on the predicted
emission-line fluxes that can lead to variations in the final results.
In addition the energies involved in the ionization potentials of the used ions are not 
high enough to identify and compare very high $T_*$  values in the
ionizing stars or clusters. The higher ionization potential is that of S$^{2+}$  so
the resulting grid is not valid to find stars hotter than 50-55 kK and the precision 
of the code at this regime is then also affected.

However, we think that this routine largely improves the accuracy and consistency
of other recipes based only on the use of emission-line fluxes in integrated ionized gaseous nebulae when
the analysis of the SED of the ionizing source cannot be carried out and it can be used as
a robust comparative tool when no evidence of a large deviation from the
simplified model exposed here is observed.

On this basis, we studied a large well-characterized sample
of \hii\ regions in the local Universe. The spectroscopic study carried out by the
{\sc chaos} collaboration has provided us with good-quality data at different galactocentric distances for the four
used emission lines and with auroral lines in many of them, which also permits a precise
determination of the gaseous ionic abundances.
The radial variation of the softness parameter in the three studied galaxies is very different, as
already observed by \citealt{pmv09}. While in NGC~628 and NGC~5457 we
observe a clear decrease of the softness parameter, in NGC~5194 the slope is positive.

Applying the {\sc HCm-Teff} code, we conclude that these radial variations can be explained in terms of
a real hardening of the incident ionizing radiation in NGC~628 and NGC~5457. On the contrary there is not
any clear evidence of a radial variation of $T_*$ in NGC~5194 despite the radial decrease of $\eta$. In none of the three
galaxies do we detect a large radial variation of $U$. No large differences in these results are found when we limit
the sample of studied \hii\ regions to those with a direct determination of the chemical abundance or to the sample
of the most luminous objects to limit the effects of stochastic 
fluctuations on the initial mass function. However, we cannot discard the influence of the radial variation of metallicity on the observed patterns of the softness parameter, as in NGC~5491, the unique object in this sample with a negative slope in $\eta$, the O/H slope is flat when the innermost \hii\ regions with no 
measurement of any auroral line are taken into account.

Using the data compiled by \cite{pmv09} to obtain more results we observe a certain trend
to find galaxies with negative slope of $\eta\prime$ in concordance with a negative gradient of 
$Z$ and a positive gradient of excitation and
$T_*$ as obtained by our code, but no clear correlation
is found for any of the studied slopes.
The existence of a global correlation between excitation or $T_*$ and metallicity is observed 
in both cases, so the influence of these two physical properties on the global
behavior of the softness parameter must be extracted.

On the other hand a certain trend of having steeper $T_*$ gradients is observed  
in less bright galaxies, which has been widely observed
in the case of gradients of $Z$.
On the contrary a large difference is not found in the case of excitation.
Therefore, there is some evidence pointing towards a real hardening of the field of radiation in spiral
discs as a consequence of the native metallicity gradient observed in many
of these objects.
In all cases, a more comprehensive study of a large sample of disc galaxies with
good determinations of the involved emission lines
necessary to break the $T_*$-$U$ degeneracy and accurate
oxygen abundances are required to give a statistically significant answer.

\section*{Acknowledgements}
We thank an anonymous referee whose very thorough and constructive comments have helped to improve this present manuscript.
This work has been partly funded by the Spanish MINECO
projects Estallidos 5
AYA2013-47742-C04 and Estallidos 6 AYA2016-79724-C4.
and the Junta de 
Andaluc\'\i a for grant EXC/2011 FQM-7058.
EPM also acknowledges support from the CSIC intramural grant 20165010-12
and the assistance from his guide dog Rocko without whose daily help this work would have been much more difficult.
RGB acknowledges support from the Spanish Ministerio de Econom\'ia y Competitividad, through 
projects AYA2016-77846-P and AYA2014- 57490-P.
We also thank Almudena Zurita and Estrella Florido for kindly providing us with the values of the
effective radii of NGC~628, NGC~5194, and NGC~5457 as measured on optical
images of these galaxies.




\bibliographystyle{mnras}
\bibliography{HCm_Teff} 








\bsp	
\label{lastpage}
\end{document}